\documentclass[fleqn,usenatbib]{mnras}

\usepackage{newtxtext,newtxmath}

\usepackage[T1]{fontenc}

\DeclareRobustCommand{\VAN}[3]{#2}
\let\VANthebibliography\thebibliography
\def\thebibliography{\DeclareRobustCommand{\VAN}[3]{##3}\VANthebibliography}

\usepackage{graphicx}	
\usepackage{amsmath}	
\usepackage{longtable}
\usepackage{multirow}
\usepackage{tablefootnote}
\usepackage{threeparttable}

\title[MeerKAT pulsar parallaxes and proper motions]{MeerKAT Pulsar Timing Array parallaxes and proper motions}

\author[M. Shamohammadi et al.]{
M.~Shamohammadi,$^{1,2}$\thanks{E-mail:msh.ph.ir@gmail.com}
M.~Bailes,$^{1,2}$
C.~Flynn,$^{1,2}$
D.~J.~Reardon,$^{1,2}$
R.~M.~Shannon,$^{1,2}$
S.~Buchner,$^{3}$\newauthor
A.~D.~Cameron,$^{1,2}$
F.~Camilo,$^{3}$
A.~Coronigu,$^{4}$
M.~Geyer,$^{3,5}$
M.~Kramer,$^{6,7}$
M.~Miles,$^{1,2}$
R.~Spiewak$^{7}$
\\
\\
$^{1}$Centre for Astrophysics and Supercomputing, Swinburne University of Technology, PO Box 218, Hawthorn, VIC 3122, Australia\\
$^{2}$ARC Centre of Excellence for Gravitational Wave Discovery (OzGrav), Mail H29, Swinburne University of Technology, PO Box 218, Hawthorn, VIC 3122, Australia\\
$^{3}$South African Radio Astronomy Observatory, 2 Fir Street, Black River Park, Observatory 7925, South Africa (SARAO)\\
$^{4}$INAF - Osservatorio Astronomico di Cagliari, Via della Scienza 5, 09047 Selargius (CA), Italy\\
$^{5}$Department of Astronomy, University of Cape Town, Rondebosch, Cape Town 7700, South Africa\\
$^{6}$Max-Planck-Institut f\"ur Radioastronomie, Auf dem H\"ugel 69, 53121 Bonn, Germany\\
$^{7}$Jodrell Bank Centre for Astrophysics, Department of Physics and Astronomy, University of Manchester, Manchester M13 9PL, UK
}

\date{Accepted XXX. Received YYY; in original form ZZZ}

\pubyear{2015}

\begin{document}
\label{firstpage}
\pagerange{\pageref{firstpage}--\pageref{lastpage}}
\maketitle

\begin{abstract}

We have determined positions, proper motions, and parallaxes of $77$ millisecond pulsars (MSPs) from $\sim3$ years of MeerKAT radio telescope observations.
Our timing and noise analyses enable us to measure $35$ significant parallaxes ($12$ of them for the first time) and $69$ significant proper motions. Eight pulsars near the ecliptic have an accurate proper motion in ecliptic longitude only. PSR~J0955$-$6150 has a good upper limit on its very small proper motion ($<$0.4 mas yr$^{-1}$). We used pulsars with accurate parallaxes to study the MSP velocities. This yields $39$  MSP transverse velocities, and combined with MSPs in the literature (excluding those in Globular Clusters) we analyse $66$ MSPs in total. We find that MSPs have, on average, much lower velocities than normal pulsars,
with a mean transverse velocity of only $78(8)$ km s$^{-1}$ (MSPs)
compared with $246(21)$ km s$^{-1}$ (normal pulsars).
We found no statistical differences between the velocity distributions of isolated and binary millisecond pulsars.
From Galactocentric cylindrical velocities of the MSPs, we derive 3-D velocity dispersions of $\sigma_{\rho}$, $\sigma_{\phi}$, $\sigma_{z}$ = $63(11)$, $48(8)$, $19(3)$ km s$^{-1}$.
We measure a mean asymmetric drift with amplitude $38(11)$ km s$^{-1}$, consistent with expectation for MSPs, given their velocity dispersions and ages.
The MSP velocity distribution is consistent with binary evolution models that predict very few MSPs with velocities $>300$ km s$^{-1}$ and a mild anticorrelation of transverse velocity with orbital period.

\end{abstract}



\section{Introduction}

Studying the transverse velocities of pulsars enables us to better understand their origins in the Milky Way, the evolutionary scenarios leading up to their birth, and the kick velocities that they receive when they are born \citep{las82,cc97}.  
Distances and velocities of pulsars also enable us to constrain the dynamical models of supernovae \citep[e.g.][]{gj95} and understand the kinematics of binary evolution \citep[e.g.][]{gsf+11,fbw+11}. 

Transverse velocities for radio pulsars are primarily determined using two different methods: firstly, measurements of pulsar proper motions and parallaxes through pulsar timing or Very Long Baseline Interferometry (VLBI) \citep[e.g.][]{rsc+21,dds+22}, and secondly, using interstellar scintillation patterns in a pulsar's dynamic spectrum \citep[e.g.][]{bpl+02,rcb+20}. Slow (or young) pulsars possess timing noise that makes
simple fits for proper motion unreliable and most
of the first proper motions determined were reliant on interferometric techniques \cite[e.g., ][] {las82,bmk+90,hla93}.

A breakthrough study of pulsar transverse velocities was made by \citet{hll+05}. They collated
all the pulsar proper motions from the literature and significantly added to them with a new timing technique from the Jodrell Bank timing data
to explore a total sample of 
$233$ pulsar proper motions. They found the two-dimensional velocities of young pulsars range from a few tens to $\approx 1600$ km s$^{-1}$. 
They used a novel deconvolution technique to derive a mean 3D pulsar birth velocity of approximately $400\pm 40$ km s$^{-1}$ for young ($<$3 Myr) pulsars, suggesting that pulsars receive large
kicks at birth.
Such high kick velocities imply that the local convective instability in the collapsed stellar core \citep{lcc01} is unlikely to be a pulsar kick mechanism (the mechanism that causes the neutron star to get kicked with a different velocity compared to its progenitor star after the supernova explosion), and they favoured more energetic mechanisms such as global asymmetric perturbations or neutrino emission in the supernovae. They found that a single Maxwellian distribution adequately fit the distribution of velocities, and derived a one-dimensional root-mean-square (rms) velocity of $265$ km s$^{-1}$, much higher than their progenitors, the OB stars which typically only possess velocities of a few tens of km s$^{-1}$.

Slowly rotating pulsars cannot be timed sufficiently accurately
to yield reliable timing parallaxes and so for many years pulsar
parallaxes were rare.
In the 2000s, dedicated very long baseline
arrays such as the Very Long Baseline Array and efficient pipelines 
made pulsar proper motion and parallax studies more routine, and several large-scale surveys were conducted on both slow-spinning pulsars and MSPs that greatly expanded our knowledge of their kinematics and distances \citep{cbv+09,dgb+19,dds+22}.
Millisecond pulsar timing arrays have also started significantly contributing
to the number of MSPs with measured timing parallaxes. PTAs now routinely
obtain sub-microsecond timing precision, and this
is permitting the measurement of many MSP parallaxes via the pulse timing method
\cite[e.g., ][]{dcl+16,abb+18,rsc+21}. In some special
cases, the non-zero apparent orbital period derivative
of pulsars enables a very model independent
estimate of their distance \cite[e.g., ][]{bb96}.

In this work, we concentrate on measurements of transverse velocities for MSPs, and substantially increase the sample with which to study their distribution. For the purposes of this paper, MSPs are defined as having a spin period of $P \leq 30$ ms and a spin-down rate of $\dot{P} \leq 10^{-17}$. 

MSPs are well known to have extremely precise spin periods and are accurate clocks. Their short spin periods and timing stability enhances the reliability of timing models. We are often able to predict and measure the arrival times of pulses (ToAs) of MSPs with  sub-microsecond precision. By having such precise ToAs, we can use the pulsar timing technique to measure the positions of MSPs often to sub-milliarcsecond accuracies. Having several years of data allows us to also measure the proper motions with the precision of better than milliarcseconds per year and even reach sub-milliarcsecond parallaxes. Using these proper motions and parallaxes, we can then derive the distances and transverse velocities with a precision of sometimes the order of a few percent and few $\rm km \, s^{-1}$, respectively \cite[e.g., ][]{mnf+16,rsc+21}. To do so, we need to observe pulsars regularly over the course of at least one year in order to correct for the R{\o}mer delay, which causes a nearly sinusoidal variation of the pulse's travel time due to the Earth's orbit around the Sun and allows us to measure the position of the pulsar. A transverse motion
of the pulsar on the sky causes a linearly increasing sinusoid with a period of one year to
appear in the timing residuals, and usually after about $3$ years
the proper motions can be determined to high accuracy.
The curvature of the pulse train's wavefront due to its finite distance of origin
allows us to measure a pulsar's parallax, provided it is not too
far from the ecliptic plane --- unfortunately an MSP at the ecliptic pole has almost no discernable parallax timing signature. One pulsar in our sample, (PSR J0711--6830) is almost 83$^\circ$ 
from the ecliptic plane and has never had its timing parallax determined.

The applications enabled by possessing pulsar parallaxes are many. For example, combining a parallax-derived distance and the pulsar's dispersion measure (as will be explained in Section \ref{sec:timing_noise}) can be used for developing Galactic electron density distribution models such as TC93 \citep{tc93}, NE2001 \citep{cl02,cl03}, and YMW16 \citep{ymw17}. These models allow us to make estimates
of pulsar distances when parallaxes are unavailable, using the DM as a proxy for distance. When combined with a proper motion, these
yield an estimate of the pulsar's transverse velocity.
In addition, the Galactic magnetic field parallel to the line of sight can be mapped out through the measurements of the distances, DMs and the Faraday rotation of pulsars \cite[e.g., ][]{hml+06,sbm+22}. Moreover, pulsar distances can be used to 
correct for 
the Shklovskii effect, which otherwise is the main contaminant to the
observed period and orbital period derivatives from their
intrinsic values \citep{shk70}.
This leads to improved estimates of the pulsar's characteristic age,
magnetic field strength and in tests of General Relativity that
require the orbital period derivative
\citep{ctk94,bb96}.

The focus of this paper is on MSP parallaxes, proper motions and
hence distances and transverse velocities. We use these to
assemble the largest-ever sample of MSPs with transverse velocities
to study how they might differ from the slowly rotating pulsars,
and also as a population.
Determining pulsar transverse velocities requires measurement of pulsar distances and parallaxes, and consequently, we have selected a sample of MSPs observed by the MeerKAT timing program \cite[MeerTime,][]{bbb+18} aimed at improving as many proper motions and parallaxes as possible.  

In Section \ref{sec:data_sets}, we explain how the MeerKAT observations were performed and how the ToAs were determined. In Section \ref{sec:methods}, we describe our methods for timing each pulsar and the modelling of timing noise, and the methods employed for the measurement of parallaxes and proper motions. In Section \ref{sec:results-measurements}, we present the positions, proper motions, and parallaxes. In Section \ref{sec:results-analysis}, we derive distances and velocities for our sample. We also present the velocity distributions and the dispersions of velocity components in Galactocentric coordinates. In Section  \ref{sec:discussions}, we compare our results to the previous work and provide discussions about the velocity distributions and the velocity dispersions. In Section \ref{sec:conclusions}, we list our conclusions and discuss the implications of our findings.

\section{The data sets}\label{sec:data_sets}
We used the first MeerKAT pulsar timing array (MPTA) data release provided by \citet{msb+22a} as the foundation of our study. This data set contains 78 MSPs observed with approximately a two week cadence with the 64-dish MeerKAT radio telescope over a period of $\sim2.5$ years, starting from early $2019$. We augmented the timing baseline of the data set to $\sim 3$ years by adding all data taken after the first data release and up to May $2022$. 

The observations were made with the L-band receiver, operating at the centre frequency of $1284$ MHz with $856$ MHz of frequency bandwidth \citep{bja+20}, and recorded with $1024$ frequency channels. Following \cite{sbm+22}, we removed the top and bottom 48 channels ($\sim 10\%$ of the total bandwidth) from the L-band receiver where response roll-off affects the signal-to-noise ratio, and the remaining $928$ channels were averaged in time and polarization to form Stokes I profiles. The typical minimum length of the observations was $4$ minutes, but depending on the brightness of the pulsars and the orbital phase of the binary pulsars, they could be up to $4$ hours long. 

Timing baselines of observations ranged from $2.33$ yr (for J1652$-$4838) to $3.26$ yr (for PSR~J1909$-$3744). We used only $1.98$ years of PSR~J1713$+$0747 observations taken prior to a change in its pulse shape between April $16$ and $17$, $2021$ \cite[e.g., ][]{jcc+22} that spoil
its timing properties.
Channels affected by radio frequency interference (RFI) were zero-weighted using MeerGuard, a modified version of CoastGuard \citep{lkg+16} for removing RFI from the MeerKAT observations. Almost no
observations were deleted due to RFI, although a small percentage
had to be removed because of observatory set-up issues such as poor phasing of the array. These
outliers were very obvious and did not present any analysis
issues.

We formed sub-banded pulse profiles by averaging frequency channels by a factor of $58$ in order to detect any frequency-dependent variation of dispersion measure (DM) across the band \citep{kcsh+13,jml+17,dvt+20,nag+22} in addition to any frequency-dependent trends that could be attributed to the intrinsic profile shape changes across frequency \citep{kll+99,amg07,xsz+21}.
This left us with 16 sub-bands that could be used to model dispersion measure variations.

We produced smoothed, high signal-to-noise, standard `template' profiles for each pulsar from the sub-banded pulse profiles for use in measuring ToAs. Each template represents a set of denoised analytical pulse profiles that evolve across frequency \citep{pdr14,p19}, using the PulsePortraiture software package\footnote{\url{https://github.com/pennucci/PulsePortraiture/tree/py3}}. Employing a 2D portrait enabled us to calculate ToAs for each sub-band, while simultaneously examining profile stability and correct for the frequency-dependent trends (if they are significant) due to time-dependent dispersion measure.

The most recently available ephemerides for the pulsars \citep[][private communication]{sbm+22} were used to initiate the timing analysis. Using the \texttt{pat} software tool in the \textsc{psrchive} package \citep{vdo12}, we measured the ToAs for all the observations, setting the threshold for timing at a signal-to-noise ratio SNR $>10$. The number of ToAs ranged from $269$ for PSR~J1327$-$0757 to $3074$ for PSR~J1909$-$3744.
We describe the timing analysis of this set of ToAs in the next section.

\section{Methods}\label{sec:methods}
\subsection{Timing and Noise analysis}\label{sec:timing_noise}
The astrometric, period, spin-down, DM, and (where necessary) binary parameters of every pulsar are described in a timing model. Using the pulsar timing software \textsc{tempo2} \citep{hem06,ehm06}, we refined the timing models by fitting for the par-file parameters. \textsc{tempo2} uses a linear least-squares approach to fitting parameter values, such that some of the parameters that have non-linear forms need to be fit multiple times in order to yield the best timing residuals. We also include systematic time jumps in our timing models
derived from the entire population. We used the JPL DE440 ephemeris as the model of the Earth's orbit around the Solar System barycentre \citep{pfw+21}, and the ToAs were referred to the TT(BIPM2020)\footnote{\url{https://webtai.bipm.org/ftp/pub/tai/ttbipm/TTBIPM.2020}}, and the Barycentric Coordinate Time (TCB) was used as the coordinate time standard for calculating the orbits in the Solar System \citep{sf92}. 

Once accurate timing models had been produced for each pulsar, they were then employed for the noise analysis using the Bayesian inference software \textsc{temponest} \citep{lah+14}. The advantage of using \textsc{temponest} in our analysis is that it uses Multinest \citep{fh08,fhb09} to sample both timing parameter and noise parameter spaces efficiently. The noise parameters were modeled by three white noise terms, an achromatic red noise term, and a dispersion noise (i.e.\ due to DM variations) term. 

\begin{itemize}
    \item White noise: there are three contributions for the white noise as follows:
    \begin{enumerate}
        \item EFAC: while calculating the ToAs described in Section \ref{sec:data_sets}, the radiometer noise is a certain contributor to the ToA uncertainties $\sigma_{\rm i}$. EFAC is modelled using a parameter, which is multiplied by the ToA uncertainties and scales them. An EFAC estimate is made for each pulsar/receiver combination. If EFAC deviates too far from unity the error estimates are not well understood.
        \item EQUAD: The EFAC parameter might not be able to fully account for the actual error, and we can broaden the errors by defining EQUAD parameter which is added in quadrature to EFAC.
        As for the EFAC parameter, an EQUAD estimate is made for each pulsar/receiver combination. Again if EQUAD is much greater
        than the rms residual, this is concerning as it means there
        are systematic errors in the arrival times that are of unknown origin.
        \item ECORR: for sub-banded data, there is likely to be a correlation between ToAs at different frequencies that are collected simultaneously because of issues such as pulse jitter \citep{abb+15}, and this introduces another noise term that can be described using ECORR. In this analysis, we used the TECORR estimator, as described in \citet[][Eq. 1]{bja+20}. ECORR is also estimated for each pulsar/receiver combination. 
    \end{enumerate}
    
    \item Red noise: this is thought to be related to the irregularities of the rotation of neutron stars \citep{sc10}. Red noise is a stochastic, achromatic term with a power-law spectrum of the form         \begin{equation}
            S_{\rm red}(f)=A_{\rm red}^{2}\left ( \frac{f}{yr^{-1}} \right )^{\gamma_{\rm red}}
        \end{equation}
    where $S_{\rm red}(f)$ is the power spectral density of red noise with fluctuation frequency $f$, $A_{\rm red}$ is the amplitude in $\mu$s yr$^{1/2}$, and $\gamma_{\rm red}$ is the spectral index.
    
    \item Dispersion noise: The ToAs are affected by frequency dispersion caused by the ionized material in the interstellar medium (ISM), and ToAs at lower observing frequencies arrive with a delay with respect to the higher ones. This delay, measured in seconds, is equal to $1/(2.41\times10^{-4})\, \textrm{DM}/\nu^{2}$, where DM is dispersion measure (in $\rm pc \, cm^{-3}$), and $\nu$ is the observing frequency (in MHz) \citep{lk05}. Due to the space motions of the Earth and pulsars, as well as the bulk motion of the ISM, DM shows a temporal variation \citep{kcsh+13,jml+17}. This variation introduces a dispersion noise term that can be modelled using a power law with the same form as the red noise. The only difference is the frequency-dependent amplitude.
\end{itemize}

For every pulsar, we performed four noise analyses with the noise models including (1) DM, red, and white noise parameters (2) DM and white noise parameters only (3) red and white noise parameters only and (4) white noise parameters only. We calculated a log Bayesian evidence for every noise model. By subtracting the log Bayesian evidence values of each of two noise models we can measure the log Bayes factor of $\ln{\mathcal{B}}$. If $\mathcal{B} > 10$ (corresponding to the log Bayes factor of $2.3$) \citep{kr95} the noise model with higher Bayesian evidence is more significant than the other and is preferred as the final noise model. 
The prior ranges for the noise parameters are listed in Table \ref{tab:priors}.

\begin{table}
\centering
\caption[Uniform prior distributions of noise parameters]{Uniform prior distributions of noise parameters are adopted in the ranges given in the second column}
\begin{tabular}{ll}
\hline
Parameter & Prior \\
\hline
$\log_{10}$[EFAC] & [$-1$,$1.5$] \\
$\log_{10}$[EQUAD] & [$-10$,$-5$] \\
$\log_{10}$[ECORR] & [$-10$,$-5$] \\
$\log_{10}$[A$_{\rm red}$] & [$-20$,$-10$] \\
$\gamma_{\rm red}$ & [$0$,$6$] \\
$\log_{10}$[A$_{\rm DM}$] & [$-20$,$-10$] \\
$\gamma_{\rm DM}$ & [$0$,$6$] \\
\hline
\end{tabular}
\label{tab:priors}
\end{table}

\subsection{Astrometry}\label{sec:astrometry}

\subsubsection{Position, Proper Motion, and Parallax Measurements}
After determining a preferred noise model, we measured the pulsar's astrometric parameters using the ToAs, and ascertained which are measured with sufficient statistical significance for use in the later tangential velocity analysis. For each pulsar the astrometric parameters are the equatorial coordinates for the position ($\alpha, \delta$), the proper motions ($\mu_\alpha\equiv\dot{\alpha}\cos\delta, \mu_{\delta}$) and the parallax. Due to the high sensitivity of MeerKAT observations, the position of pulsars can be obtained extremely well, and we did not need to sample for the position parameter; so, these remained fixed. We performed three analyses using \textsc{temponest} sampling for (1) proper motion, (2) parallax, and (3) proper motion and parallax.
Bayes factors produced by \textsc{temponest} showed that, for all pulsars, the proper motion and parallax were detectable (with $\mathcal{B} > 10$) and superior to models without them. In addition, we used ecliptic coordinates for the position ($\lambda, \beta$) and proper motion ($\mu_\lambda\equiv\dot{\lambda}\cos\beta, \mu_{\rm \beta}$), and repeated the analyses and compared the Bayesian evidence for these solutions with the equatorial ones, and found them to be similar. Note that covariance between position and proper motion parameters was minimized by setting the epoch of the positions to be near the middle of the data set's baseline in time. \citet{mnf+16} demonstrated that positional uncertainties reported in ecliptic coordinates are less than the equatorial coordinates. Therefore, we used timing analyses in ecliptic coordinates to derive the Galactic position ($l, b$) and Galactic proper motion ($\mu_{l}\equiv\dot{l}\cos b, \mu_{b}$) of all pulsars using the Astropy package \citep{astropy}.

\subsection{Distances}

\subsubsection{Distances from parallax measurement}
The most direct method for measuring distance is parallax $d = \varpi^{-1}$, where $\varpi$ is trigonometric parallax in mas and $d$ is in kpc. This method does not depend on any intrinsic physical property of the object and is purely geometric, and thus model-independent. For pulsars, the two most common methods of measuring parallax are using pulsar timing solutions and VLBI observations. As will be seen, these two methods dominate the distances used in section \ref{sec:results-analysis} for our transverse velocity analysis.   

\subsection{Velocities}
The full 3D velocity of a celestial object relative to the Sun can be broken into two components: a radial velocity component and a transverse (or tangential) velocity component. The radial velocity for stars in the Milky Way is typically estimated using the Doppler effect, although this is not possible for pulsars, as no spectral lines are available unless
they are in a binary system with a star. The transverse velocity (in km s$^{-1}$), is the velocity perpendicular to the line of sight and is given by  
\begin{equation}
    v_\perp = 4.74\,{\rm km\,s}^{-1} \left( \frac{\mu}{\rm mas\,yr^{-1}}\right) \left( \frac{d}{\rm kpc} \right),
\end{equation}
where $\mu$ is the total proper motion, and $d$ is the distance \citep{lbh97}.

\subsubsection{Correcting for the Local Standard of Rest and the Galactic differential rotation}\label{sec:method_vel_corr}
Proper motions are usually measured relative to the 
Solar System's barycentre, but if we want to understand their
origins as Milky Way objects, we need to adjust for the reference frame. To obtain the velocity of pulsars with respect to the Galactic Center (GC), we need to correct the relative proper motions for two effects: (1) the peculiar velocity of the Sun with respect to the
local standard of rest (LSR) and (2) differential rotation of the Galaxy \citep{hll+05}. There are two types of LSR, both of which are defined as reference frames at the location of the Sun. The first is the `kinematic LSR' which orbits the Galaxy with the mean velocity of the neighbouring stars ($\bar{v}_{\rm \phi}$). The second is the `dynamic LSR' which rotates around the Galaxy in a perfectly circular orbit with a velocity that is exerted by the gravitational potential of the Galaxy ($v_{\rm c}$). The difference between the two ($v_{\rm c} - \bar{v}_{\rm \phi}$) is defined as the ``asymmetric drift'' \citep{gjb+13,fer19}.

The Galactic disk rotates differentially, and the Galactic orbital periods of stars are dependent on their distances from the GC. The circular rotation speed of stars at distance $R$ from the GC defines the rotation curve of the Galaxy. In the last decade, there have been a number of studies to improve the measured rotation curve of the Galaxy using a variety of different methods \cite[e.g., ][]{bab+12,bck14,rmb+14,rmb+19}. In this work, we used the results and rotation curve model from \citet{rmb+19}.

For every MSP, we transformed the coordinates of the pulsars from their ecliptic coordinates to the cartesian Galactocentric coordinates ($X$, $Y$, $Z$) using Astropy\footnote{\url{https://docs.astropy.org/en/stable/api/astropy.coordinates.Galactocentric.html}}. The distance from the Sun to the Galactic mid-plane was taken from \citet{rmb+19} ($Z_{\odot} = 5.5(58)$ pc), and the distance from the Sun to the GC was taken from \citet{aaa+18} ($R_{\odot} = 8.12(3)$ kpc). We then derived the Galactocentric velocity components ($V_{\rm X}$, $V_{\rm Y}$, $V_{\rm Z}$). For subtracting the Sun's peculiar velocity from these components, we used the values of ($U$, $V$, $W$)$_{\odot}$ = ($11.10(74), 12.24(47), 7.25(37)$) km s$^{-1}$ from \citet{sbd10}. 

To correct for differential rotation of the Galaxy, we calculated the distances of the pulsars from the Galactic centre, and used the universal rotation curve of the Galaxy from \citet[][Tab. 4]{rmb+19} to calculate the $X$ and $Y$ components of the Galactic rotation at the position of MSPs. Finally, we subtracted these from the velocities of MSPs and derived their transverse velocities with respect to their own LSR. All velocities reported in this work are thus
corrected for the peculiar velocity of the Sun and the rotation curve of the Galaxy. 


\section{Results: astrometry}\label{sec:results-measurements}

In this section we report our astrometric results, including positions, proper motions, parallaxes for our sample pulsars, and compare with MSPs for which these have been previously measured in the literature.   

\subsection{Positions and proper motions}
In Table \ref{tab:positions}, positions from timing analyses in equatorial and ecliptic coordinates for $77$ MSPs are given. These were not sampled using \textsc{temponest} and only fitted using \textsc{tempo2}. In addition, the positions in Galactic coordinates are derived from ecliptic coordinates using Astropy. All uncertainties shown are computed via the median with $68.3\%$ equally-tailed credible intervals.

In Table \ref{tab:proper_motions}, we show proper motions from timing analyses in both equatorial and ecliptic coordinates. In this table, the proper motions in Galactic coordinates are derived from the ecliptic proper motions using Astropy. The reported values and uncertainties are the lower bound ($16\%$), median ($50\%$), and upper bound ($84\%$) of marginalized posterior probability distributions. The timing baseline of observations for each pulsar, as well as the best previous measurements of proper motions in equatorial coordinates and their corresponding references, are listed. In our study, our limited timing baseline means that our proper motion results are not superior to all previous results,
mainly because the relative error is proportional to the observing span $T^{-3/2}$, and some of the pulsars
have been timed for over a decade with other instruments. 

Out of $77$ pulsars, $69$ had proper motions with statistical significance greater than $3\sigma$, and are shown in the upper portion of Table \ref{tab:proper_motions}. The remaining $8$ pulsars (i.e.\ those listed in the lower portion of the Table) consist of an isolated MSP (J1721$-$2457) with relatively poor rms timing residuals of $3.7 \mu$s and
less than two deg from the ecliptic, and seven binary MSPs (J0955$-$6150, J1022$+$1001, J1327$-$0755, J1653$-$2054, J1705$-$1903, J1802$-$2124, J1811$-$2405) that have either poor timing residuals or small proper motions (as low as $0.21(15)$ mas yr$^{-1}$ for J0955$-$6150). 
PSR~J1022$+$1001 is near the ecliptic plane, and as expected had a very accurate proper
motion in ecliptic longitude, but a poor constraint in ecliptic latitude.
For this pulsar, we set the proper motion and orbital parameters from the high accuracy VLBI study of
\cite{dvk+16} and derived a parallax of $1.2(4)$ mas.

We typically require at least two years of data in order to separate the annual parallax from the proper motion of a pulsar, due to the motion of the pulsar against the background of the sky and the orbit of the Earth around the Sun. We note that it is not always simple to break this degeneracy for the pulsars with high rms timing residuals, and worse for pulsars in (wide) binary systems. A solution to this is to implement VLBI techniques, which can often measure pulsar parallaxes significantly better, e.g.\ \citet{dvt+08} measured the parallax for PSR~J0437$-$4715 to be $6.396(54)$ mas with the Australian Long Baseline Array.

All but two of the proper motions listed in Table \ref{tab:proper_motions} are consistent with previous results within $3\sigma$. One component of the proper motion of PSRs~J1446$-$4701 is just
over $3\sigma$ inconsistent with the values in \citet{rsc+21}, but the total
proper motion is the same to within 8\%.
The origin of this discrepancy is unclear. We checked the use of the Solar System ephemeris that \citet{rsc+21} used for their timing analysis, and our parallax remained the same
to within errors. This inconsistency might be due to the different noise models that were adopted or be due to different noise properties at different epochs.

\begin{figure*}
\centering
 \includegraphics[width=\textwidth]{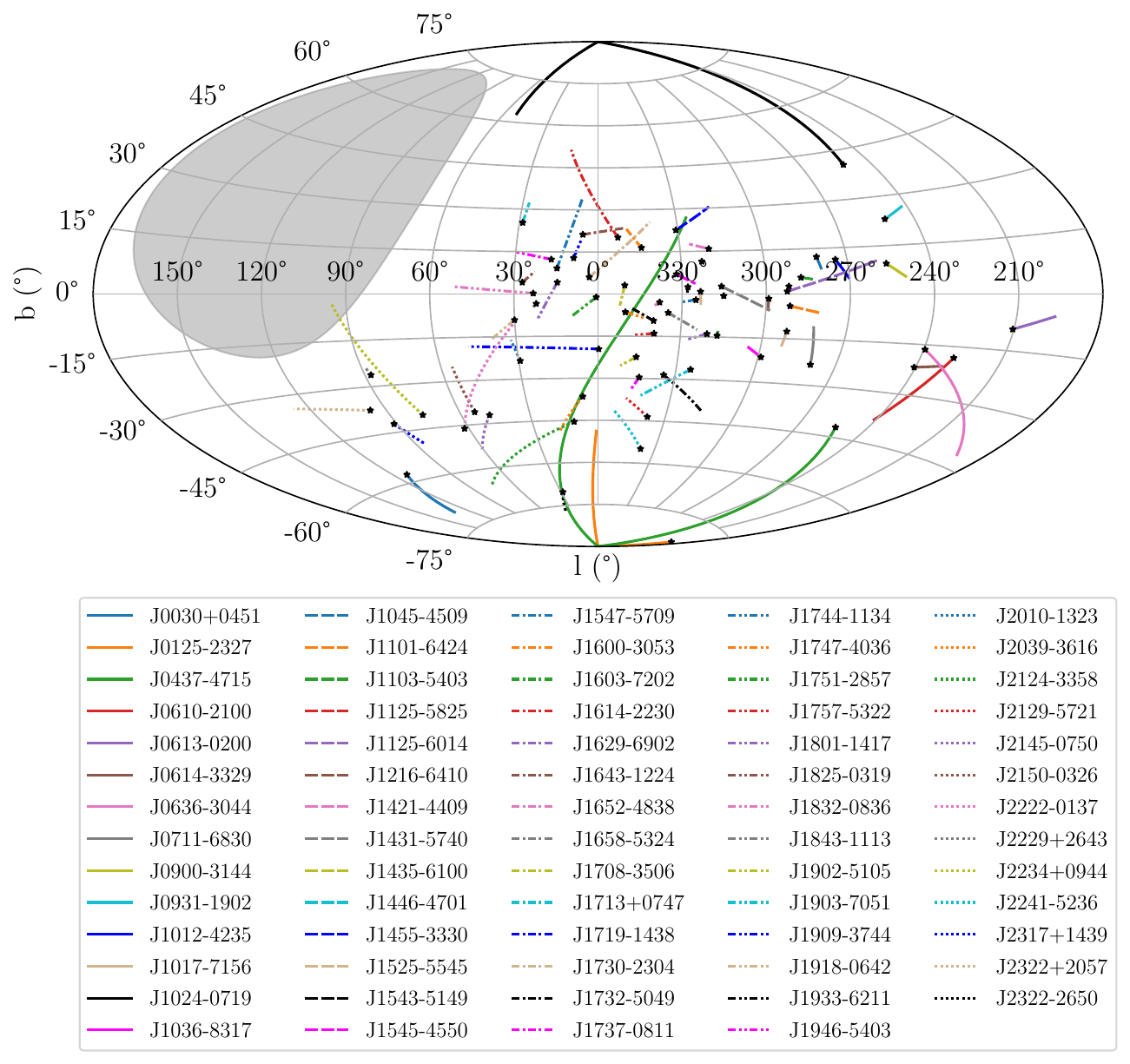}
\caption[MSP trajectories from $5$ Myr ago up to the present]{Celestial distribution of the 69 MSPs in this study, shown in Galactic longitude, $l$ and Galactic latitude, $b$. The curves are the paths of these MSPs from $5 \, \rm Myr$ up until the present, and the current positions of these MSPs are shown with black stars. The effects of the peculiar motion of the Sun and the Galactic rotation curve have been subtracted from the observed proper motions. 
The black unit vectors are showing the proper motions with at least $3\sigma$ significance, and the length of the brown vectors are showing the magnitudes of velocities with at least $3\sigma$ significance in parallax detections. See the text for more details.
}
\label{fig:aitoff_projection}
\end{figure*}

Figure \ref{fig:aitoff_projection} presents the trajectories of the $69$ MSPs since $5$ Myr ago until the present in Galactic longitude ($l$) and Galactic latitude ($b$), assuming zero radial velocity for the MSPs. Both effects of the peculiar velocity of the Sun and the Galactic differential rotation are subtracted from the observed proper motions of MSPs. In addition, the Galactic gravitational potential is assumed to be zero, because the oscillation period
in the galactic potential ($> 100 $Myr) is much larger than 5 Myr,
and so the MSPs are simply moving along their corrected proper motion vectors. 
The area in gray indicates the part of the sky that the MeerKAT radio telescope is not able to reach. From
the diagram it is obvious that the MSPs are both moving towards and away from the galactic plane, 
consistent with a relaxed ancient population.

\begin{table*}
\centering
\caption[Pulsar positions]{Pulsar positions in Equatorial ($\alpha$,$\delta$), Ecliptic ($\lambda$,$\beta$), and Galactic ($l$,$b$) coordinates to 3 significant figures. 
The digits in parentheses are showing $1\sigma$ uncertainty in the last significant digits. The last column shows the reference epochs (in MJD) at which the positions are referred to.}
\begin{tabular}{llllllllll}
\hline
\multicolumn{1}{c}{} & \multicolumn{2}{c}{Equatorial coordinates} & \multicolumn{1}{c}{} & \multicolumn{2}{c}{Ecliptic coordinates} & \multicolumn{1}{c}{} & \multicolumn{2}{c}{Galactic coordinates} & \multicolumn{1}{c}{}\\
\cline{2-3}\cline{5-6}\cline{8-9}
Pulsar & $\alpha$ & $\delta$ & & $\lambda$ & $\beta$ & & $l$ & $b$ & Epoch \\
 & (hh:mm:ss) & (dd:mm:ss) & & ($^\circ$) & ($^\circ$) & & ($^\circ$) & ($^\circ$) & (MJD) \\
\hline
J0030$+$0451 & $00$:$30$:$27.42348(5)$ & $+04$:$51$:$39.7144(17)$ & & $8.910338462(13)$ & $1.4457050(5)$ & & $113.141$ & $-57.611$ & $59251$ \\
J0125$-$2327 & $01$:$25$:$1.075151(3)$ & $-23$:$27$:$8.13167(5)$ & & $9.743403184(9)$ & $-29.830147598(15)$ & & $188.932$ & $-81.569$ & $59155$ \\
J0437$-$4715 & $04$:$37$:$16.0492624(13)$ & $-47$:$15$:$10.027930(15)$ & & $50.469059178(11)$ & $-67.873319464(5)$ & & $253.395$ & $-41.963$ & $59187$ \\
J0610$-$2100 & $06$:$10$:$13.602953(6)$ & $-21$:$00$:$27.74435(13)$ & & $93.34234093(3)$ & $-44.41732764(4)$ & & $227.746$ & $-18.184$ & $59152$ \\
J0613$-$0200 & $06$:$13$:$43.977070(3)$ & $-02$:$00$:$47.34294(13)$ & & $93.799014159(15)$ & $-25.40716602(4)$ & & $210.413$ & $-9.305$ & $59152$ \\
J0614$-$3329 & $06$:$14$:$10.348230(3)$ & $-33$:$29$:$54.13156(4)$ & & $95.410779861(16)$ & $-56.871120173(12)$ & & $240.501$ & $-21.827$ & $59155$ \\
J0636$-$3044 & $06$:$36$:$34.398358(5)$ & $-30$:$44$:$1.10761(7)$ & & $103.35534034(3)$ & $-53.74983738(2)$ & & $239.526$ & $-16.416$ & $59163$ \\
J0711$-$6830 & $07$:$11$:$54.153093(12)$ & $-68$:$30$:$47.23387(10)$ & & $204.06089121(19)$ & $-82.88867768(2)$ & & $279.531$ & $-23.280$ & $59149$ \\
J0900$-$3144 & $09$:$00$:$43.95220(3)$ & $-31$:$44$:$30.8722(4)$ & & $150.54478942(11)$ & $-46.14670208(11)$ & & $256.162$ & $9.486$ & $59149$ \\
J0931$-$1902 & $09$:$31$:$19.116246(6)$ & $-19$:$02$:$55.05445(13)$ & & $152.37696462(2)$ & $-31.77673163(4)$ & & $250.999$ & $23.054$ & $59149$ \\
J0955$-$6150 & $09$:$55$:$20.847467(18)$ & $-61$:$50$:$16.89532(13)$ & & $197.40608543(8)$ & $-64.96064654(4)$ & & $283.685$ & $-5.737$ & $59138$ \\
J1012$-$4235 & $10$:$12$:$12.938801(10)$ & $-42$:$35$:$53.40253(12)$ & & $176.78918824(4)$ & $-48.91085447(4)$ & & $274.218$ & $11.225$ & $59152$ \\
J1017$-$7156 & $10$:$17$:$51.311783(15)$ & $-71$:$56$:$41.57398(6)$ & & $222.42774900(4)$ & $-67.736237859(16)$ & & $291.558$ & $-12.553$ & $59132$ \\
J1022$+$1001 & $10$:$22$:$57.990(2)$ & $+10$:$01$:$52.92(8)$ & & $153.86581653(3)$ & $-0.06391(2)$ & & $231.794$ & $51.101$ & $59166$ \\
J1024$-$0719 & $10$:$24$:$38.648440(3)$ & $-07$:$19$:$19.98201(9)$ & & $160.734318047(8)$ & $-16.04485601(3)$ & & $251.702$ & $40.515$ & $59152$ \\
J1036$-$8317 & $10$:$36$:$40.59408(3)$ & $-83$:$17$:$56.14254(4)$ & & $253.01178441(3)$ & $-68.082219349(15)$ & & $298.943$ & $-21.497$ & $59187$ \\
J1045$-$4509 & $10$:$45$:$50.179671(15)$ & $-45$:$09$:$54.05686(17)$ & & $186.51850583(6)$ & $-47.71477807(5)$ & & $280.851$ & $12.254$ & $59142$ \\
J1101$-$6424 & $11$:$01$:$37.19154(3)$ & $-64$:$24$:$39.33316(16)$ & & $211.77519692(9)$ & $-60.54978942(5)$ & & $291.417$ & $-4.023$ & $59142$ \\
J1103$-$5403 & $11$:$03$:$33.27734(4)$ & $-54$:$03$:$43.1296(4)$ & & $198.52939001(16)$ & $-53.10724535(11)$ & & $287.421$ & $5.529$ & $59149$ \\
J1125$-$5825 & $11$:$25$:$44.352943(14)$ & $-58$:$25$:$16.85965(11)$ & & $207.30413917(5)$ & $-54.35535501(3)$ & & $291.893$ & $2.602$ & $59132$ \\
J1125$-$6014 & $11$:$25$:$55.243299(4)$ & $-60$:$14$:$6.80942(3)$ & & $209.504093810(12)$ & $-55.659051349(9)$ & & $292.504$ & $0.894$ & $59149$ \\
J1216$-$6410 & $12$:$16$:$7.321183(7)$ & $-64$:$10$:$9.16656(4)$ & & $221.617852389(19)$ & $-54.452089781(13)$ & & $299.096$ & $-1.561$ & $59132$ \\
J1327$-$0755 & $13$:$27$:$57.58052(9)$ & $-07$:$55$:$29.918(3)$ & & $203.27816789(2)$ & $1.2046411(10)$ & & $318.383$ & $53.848$ & $59281$ \\
J1421$-$4409 & $14$:$21$:$20.960491(9)$ & $-44$:$09$:$4.57445(14)$ & & $228.33837111(2)$ & $-28.29432179(4)$ & & $319.497$ & $15.809$ & $59142$ \\
J1431$-$5740 & $14$:$31$:$3.49603(4)$ & $-57$:$40$:$11.6601(4)$ & & $236.41161578(9)$ & $-40.16383928(11)$ & & $315.962$ & $2.660$ & $59129$ \\
J1435$-$6100 & $14$:$35$:$20.26018(2)$ & $-61$:$00$:$58.0019(2)$ & & $238.94141530(6)$ & $-42.97710038(6)$ & & $315.186$ & $-0.641$ & $59132$ \\
J1446$-$4701 & $14$:$46$:$35.709721(8)$ & $-47$:$01$:$26.79678(14)$ & & $234.21533611(2)$ & $-29.41052857(4)$ & & $322.500$ & $11.425$ & $59129$ \\
J1455$-$3330 & $14$:$55$:$47.976991(8)$ & $-33$:$30$:$46.40280(19)$ & & $231.347559978(19)$ & $-16.04479743(6)$ & & $330.722$ & $22.562$ & $59149$ \\
J1525$-$5545 & $15$:$25$:$28.12719(4)$ & $-55$:$45$:$49.8854(6)$ & & $244.36826552(9)$ & $-35.71040433(14)$ & & $323.439$ & $0.851$ & $59149$ \\
J1543$-$5149 & $15$:$43$:$44.145289(15)$ & $-51$:$49$:$54.7135(2)$ & & $246.13439909(4)$ & $-31.17770143(6)$ & & $327.921$ & $2.479$ & $59132$ \\
J1545$-$4550 & $15$:$45$:$55.945403(5)$ & $-45$:$50$:$37.50813(11)$ & & $244.821544044(13)$ & $-25.29111285(3)$ & & $331.892$ & $6.988$ & $59142$ \\
J1547$-$5709 & $15$:$47$:$24.121775(16)$ & $-57$:$09$:$17.5931(2)$ & & $248.44532730(4)$ & $-36.17005240(6)$ & & $325.076$ & $-2.052$ & $59162$ \\
J1600$-$3053 & $16$:$00$:$51.902514(4)$ & $-30$:$53$:$49.45343(17)$ & & $244.347678961(9)$ & $-10.07185627(5)$ & & $344.090$ & $16.451$ & $59138$ \\
J1603$-$7202 & $16$:$03$:$35.670781(16)$ & $-72$:$02$:$32.82328(9)$ & & $256.52028909(3)$ & $-49.96303799(2)$ & & $316.630$ & $-14.496$ & $59127$ \\
J1614$-$2230 & $16$:$14$:$36.50974(3)$ & $-22$:$30$:$31.542(2)$ & & $245.788315625(16)$ & $-1.2568780(7)$ & & $352.636$ & $20.192$ & $59146$ \\
J1629$-$6902 & $16$:$29$:$8.744211(5)$ & $-69$:$02$:$45.46922(3)$ & & $258.424015436(10)$ & $-46.517655909(9)$ & & $320.371$ & $-13.926$ & $59127$ \\
J1643$-$1224 & $16$:$43$:$38.16612(2)$ & $-12$:$24$:$58.6334(14)$ & & $251.08723617(7)$ & $9.7783440(4)$ & & $5.670$ & $21.218$ & $59146$ \\
J1652$-$4838 & $16$:$52$:$37.623911(18)$ & $-48$:$38$:$51.7486(4)$ & & $257.71020795(6)$ & $-25.92450458(11)$ & & $338.005$ & $-2.920$ & $59021$ \\
J1653$-$2054 & $16$:$53$:$31.01268(7)$ & $-20$:$54$:$54.977(9)$ & & $254.49648386(6)$ & $1.633926(2)$ & & $359.969$ & $14.261$ & $59149$ \\
J1658$-$5324 & $16$:$58$:$39.343556(7)$ & $-53$:$24$:$6.97315(11)$ & & $259.45266028(2)$ & $-30.52273666(3)$ & & $334.869$ & $-6.625$ & $59188$ \\
J1705$-$1903 & $17$:$05$:$43.84864(4)$ & $-19$:$03$:$41.414(5)$ & & $257.16132332(7)$ & $3.7751398(10)$ & & $3.248$ & $13.030$ & $59198$ \\
J1708$-$3506 & $17$:$08$:$17.616678(11)$ & $-35$:$06$:$22.6595(5)$ & & $259.21049750(4)$ & $-12.14991994(12)$ & & $350.470$ & $3.124$ & $59215$ \\
J1713$+$0747 & $17$:$13$:$49.5367813(12)$ & $+07$:$47$:$37.45024(4)$ & & $256.668711830(6)$ & $30.700351152(10)$ & & $28.751$ & $25.223$ & $58957$ \\
J1719$-$1438 & $17$:$19$:$10.076221(4)$ & $-14$:$38$:$1.0020(4)$ & & $260.016965101(16)$ & $8.45261155(11)$ & & $8.859$ & $12.838$ & $59150$ \\
J1721$-$2457 & $17$:$21$:$5.50008(3)$ & $-24$:$57$:$6.226(6)$ & & $261.18398611(6)$ & $-1.8096426(17)$ & & $0.387$ & $6.751$ & $59205$ \\
J1730$-$2304 & $17$:$30$:$21.68471(6)$ & $-23$:$04$:$31.259(16)$ & & $263.186095246(13)$ & $0.188838(4)$ & & $3.137$ & $6.023$ & $59150$ \\
J1732$-$5049 & $17$:$32$:$47.766068(7)$ & $-50$:$49$:$0.31465(13)$ & & $265.16176864(2)$ & $-27.49163468(4)$ & & $340.029$ & $-9.454$ & $59115$ \\
J1737$-$0811 & $17$:$37$:$47.109064(19)$ & $-08$:$11$:$8.9798(11)$ & & $264.30464763(7)$ & $15.1437208(2)$ & & $16.932$ & $12.318$ & $59150$ \\
J1744$-$1134 & $17$:$44$:$29.4220844(14)$ & $-11$:$34$:$54.80132(10)$ & & $266.119458498(6)$ & $11.80517508(3)$ & & $14.794$ & $9.180$ & $59150$ \\
J1747$-$4036 & $17$:$47$:$48.715690(13)$ & $-40$:$36$:$54.7995(5)$ & & $267.57913114(4)$ & $-17.20154481(13)$ & & $350.208$ & $-6.412$ & $59130$ \\
J1751$-$2857 & $17$:$51$:$32.686848(5)$ & $-28$:$57$:$46.5713(7)$ & & $268.141987317(19)$ & $-5.5372894(2)$ & & $0.646$ & $-1.124$ & $59150$ \\
J1757$-$5322 & $17$:$57$:$15.154960(7)$ & $-53$:$22$:$26.59604(12)$ & & $269.52715024(2)$ & $-29.93590243(4)$ & & $339.637$ & $-13.980$ & $59163$ \\
J1801$-$1417 & $18$:$01$:$51.064902(5)$ & $-14$:$17$:$34.5498(6)$ & & $270.45421998(2)$ & $9.14561928(15)$ & & $14.546$ & $4.162$ & $59150$ \\
J1802$-$2124 & $18$:$02$:$5.334393(11)$ & $-21$:$24$:$3.676(5)$ & & $270.48652542(4)$ & $2.0373763(13)$ & & $8.382$ & $0.611$ & $59127$ \\
J1811$-$2405 & $18$:$11$:$19.854352(6)$ & $-24$:$05$:$18.428(4)$ & & $272.586046336(10)$ & $-0.6745983(11)$ & & $7.073$ & $-2.559$ & $59145$ \\
J1825$-$0319 & $18$:$25$:$55.954707(11)$ & $-03$:$19$:$57.5837(5)$ & & $276.88733392(5)$ & $19.95175171(14)$ & & $27.045$ & $4.138$ & $59155$ \\
J1832$-$0836 & $18$:$32$:$27.589497(5)$ & $-08$:$36$:$55.1454(3)$ & & $278.29199042(2)$ & $14.59068359(9)$ & & $23.109$ & $0.256$ & $59147$ \\
J1843$-$1113 & $18$:$43$:$41.260405(4)$ & $-11$:$13$:$31.1031(3)$ & & $280.944529539(16)$ & $11.79998448(8)$ & & $22.055$ & $-3.397$ & $59150$ \\
J1902$-$5105 & $19$:$02$:$2.844154(10)$ & $-51$:$05$:$57.0563(2)$ & & $280.99087040(3)$ & $-28.25008697(5)$ & & $345.650$ & $-22.379$ & $59152$ \\
J1903$-$7051 & $19$:$03$:$38.783931(19)$ & $-70$:$51$:$43.58175(12)$ & & $277.69661567(4)$ & $-47.84766764(3)$ & & $324.391$ & $-26.508$ & $59136$ \\
\hline
\end{tabular}
\label{tab:positions}
\end{table*}

\begin{table*}
\centering
\contcaption{}
\begin{tabular}{llllllllll}
\hline
\multicolumn{1}{c}{} & \multicolumn{2}{c}{Equatorial coordinates} & \multicolumn{1}{c}{} & \multicolumn{2}{c}{Ecliptic coordinates} & \multicolumn{1}{c}{} & \multicolumn{2}{c}{Galactic coordinates} & \multicolumn{1}{c}{}\\
\cline{2-3}\cline{5-6}\cline{8-9}
Pulsar & $\alpha$ & $\delta$ & & $\lambda$ & $\beta$ & & $l$ & $b$ & Epoch \\
 & (hh:mm:ss) & (dd:mm:ss) & & ($^\circ$) & ($^\circ$) & & ($^\circ$) & ($^\circ$) & (MJD) \\
\hline
J1909$-$3744 & $19$:$09$:$47.4245244(7)$ & $-37$:$44$:$14.91930(3)$ & & $284.220813110(2)$ & $-15.155611509(9)$ & & $359.731$ & $-19.596$ & $59120$ \\
J1918$-$0642 & $19$:$18$:$48.027685(2)$ & $-06$:$42$:$34.95735(12)$ & & $290.314613645(9)$ & $15.35104745(3)$ & & $30.027$ & $-9.123$ & $59155$ \\
J1933$-$6211 & $19$:$33$:$32.413330(4)$ & $-62$:$11$:$46.69324(5)$ & & $283.961757663(10)$ & $-39.885817735(13)$ & & $334.431$ & $-28.631$ & $59132$ \\
J1946$-$5403 & $19$:$46$:$34.4933990(19)$ & $-54$:$03$:$42.55867(3)$ & & $288.139717466(5)$ & $-32.287682135(10)$ & & $343.885$ & $-29.580$ & $59166$ \\
J2010$-$1323 & $20$:$10$:$45.922635(4)$ & $-13$:$23$:$56.1327(3)$ & & $301.924490472(7)$ & $6.49093303(7)$ & & $29.446$ & $-23.540$ & $59155$ \\
J2039$-$3616 & $20$:$39$:$16.580827(4)$ & $-36$:$16$:$17.23958(11)$ & & $302.723254857(9)$ & $-17.24602148(3)$ & & $6.331$ & $-36.517$ & $59166$ \\
J2124$-$3358 & $21$:$24$:$43.834930(4)$ & $-33$:$58$:$45.49195(11)$ & & $312.738766685(11)$ & $-17.81894426(3)$ & & $10.925$ & $-45.438$ & $59155$ \\
J2129$-$5721 & $21$:$29$:$22.781548(3)$ & $-57$:$21$:$14.33356(4)$ & & $303.827979841(7)$ & $-39.900000443(10)$ & & $338.005$ & $-43.570$ & $59145$ \\
J2145$-$0750 & $21$:$45$:$50.453394(17)$ & $-07$:$50$:$18.5938(7)$ & & $326.02458223(2)$ & $5.3130387(2)$ & & $47.777$ & $-42.084$ & $59155$ \\
J2150$-$0326 & $21$:$50$:$27.235330(10)$ & $-03$:$26$:$32.8318(4)$ & & $328.604922236(17)$ & $9.06674597(12)$ & & $53.573$ & $-40.757$ & $59199$ \\
J2222$-$0137 & $22$:$22$:$5.996923(12)$ & $-01$:$37$:$15.7776(5)$ & & $336.73199735(2)$ & $7.97708619(14)$ & & $62.019$ & $-46.075$ & $59155$ \\
J2229$+$2643 & $22$:$29$:$50.883815(5)$ & $+26$:$43$:$57.61393(13)$ & & $350.695629791(19)$ & $33.29016684(4)$ & & $87.693$ & $-26.284$ & $59251$ \\
J2234$+$0944 & $22$:$34$:$46.856913(9)$ & $+09$:$44$:$30.0415(3)$ & & $344.11900953(3)$ & $17.31853087(9)$ & & $76.280$ & $-40.438$ & $59251$ \\
J2241$-$5236 & $22$:$41$:$42.0400897(10)$ & $-52$:$36$:$36.272400(11)$ & & $318.696416363(3)$ & $-40.393437599(3)$ & & $337.457$ & $-54.927$ & $59132$ \\
J2317$+$1439 & $23$:$17$:$9.235606(4)$ & $+14$:$39$:$31.29559(12)$ & & $356.129406224(12)$ & $17.68024246(4)$ & & $91.361$ & $-42.360$ & $59251$ \\
J2322$+$2057 & $23$:$22$:$22.320041(6)$ & $+20$:$57$:$2.50080(14)$ & & $0.135924003(19)$ & $22.87835838(4)$ & & $96.515$ & $-37.310$ & $59251$ \\
J2322$-$2650 & $23$:$22$:$34.638615(5)$ & $-26$:$50$:$58.38398(10)$ & & $340.443069768(14)$ & $-20.89690479(3)$ & & $28.637$ & $-70.228$ & $59155$ \\
\hline
\end{tabular}
\end{table*}

\begin{table*}
\centering
\caption[Pulsar proper motions]{Pulsar Proper Motions in Equatorial, Ecliptic, and approximate Galactic Coordinates. The Galactic proper motions are derived from the Ecliptic ones. The time span (column 2) of observations for each pulsar is provided in yr. The previous best measurements of proper motions in Equatorial coordinates and the corresponding references are shown in the final 3 columns. Proper motions with $>3\sigma$ and $<3\sigma$ detections are listed in the upper and lower portions of the table, respectively. The list of the references are provided at the end of the table.}
\hspace*{-1.5cm}
\begin{threeparttable}
\begin{tabular}{@{}llllllllllllll}
\hline
\multicolumn{1}{c}{} & \multicolumn{1}{c}{} & \multicolumn{2}{c}{Equatorial coordinates} & \multicolumn{1}{c}{} & \multicolumn{2}{c}{Ecliptic coordinates} & \multicolumn{1}{c}{} & \multicolumn{2}{c}{Galactic coordinates} & \multicolumn{1}{c}{} & \multicolumn{3}{c}{Best Previous Measurement} \\
\cline{3-4}\cline{6-7}\cline{9-10}\cline{12-14}

Pulsar & Span & $\mu_\alpha\equiv\dot{\alpha}\cos\delta$ & $\mu_{\delta}$ & & $\mu_\lambda\equiv\dot{\lambda}\cos\beta$ & $\mu_{\beta}$ & & $\mu_{l}\equiv\dot{l}\cos b$ & $\mu_{b}$ & & $\mu_\alpha$ & $\mu_{\delta}$ & Ref \\
 & (yr) & (mas\,yr$^{-1}$) & (mas\,yr$^{-1}$) & & (mas\,yr$^{-1}$) & (mas\,yr$^{-1}$) & & (mas\,yr$^{-1}$) & (mas\,yr$^{-1}$) & & (mas\,yr$^{-1}$) & (mas\,yr$^{-1}$) &  \\
\hline
\multicolumn{14}{c}{Total Proper Motion Detections $>3\sigma$}\\
\hline
J0030$+$0451 & $2.54$ & $-6.4(11)$ & $1(2)$ & & $-5.51(7)$ & $3(3)$ & & $-6.2(6)$ & $2(3)$ & & $-6.2(1)$ & $0.5(3)$ & (1) \\
J0125$-$2327 & $3.06$ & $37.15(5)$ & $10.75(6)$ & & $38.18(3)$ & $-6.17(6)$ & & $7.70(6)$ & $37.90(4)$ & & & & \\   
J0437$-$4715 & $2.86$ & $121.394(19)$ & $-71.489(19)$ & & $85.96(2)$ & $-111.60(2)$ & & $62.89(2)$ & $126.06(2)$ & & $121.4385(20)$ & $-71.4754(20)$ & (2) \\
J0610$-$2100 & $3.05$ & $9.29(11)$ & $16.74(17)$ & & $8.86(11)$ & $16.97(17)$ & & $-11.81(17)$ & $15.08(15)$ & & $9.11(3)$ & $16.45(3)$ & (3) \\
J0613$-$0200 & $3.05$ & $1.81(6)$ & $-10.31(17)$ & & $2.08(6)$ & $-10.28(17)$ & & $10.00(15)$ & $-3.11(11)$ & & $1.836(5)$ & $-10.349(13)$ & (6) \\
J0614$-$3329 & $3.06$ & $0.60(4)$ & $-1.84(5)$ & & $0.68(4)$ & $-1.81(5)$ & & $1.93(4)$ & $-0.02(4)$ & & $0.61(3)$ & $-1.74(4)$ & (4) \\
J0636$-$3044 & $2.99$ & $26.01(9)$ & $-19.26(10)$ & & $27.92(9)$ & $-16.37(10)$ & & $27.66(8)$ & $16.79(11)$ & & & & \\   
J0711$-$6830 & $3.06$ & $-15.54(5)$ & $14.24(6)$ & & $-12.11(5)$ & $-17.25(5)$ & & $-17.79(6)$ & $-11.21(5)$ & & $-15.568(10)$ & $14.175(11)$ & (6) \\
J0900$-$3144 & $3.06$ & $-1.4(4)$ & $2.4(5)$ & & $-2.3(3)$ & $1.6(5)$ & & $-2.7(4)$ & $0.5(6)$ & & $-1.01(5)$ & $2.02(7)$ & (5) \\
J0931$-$1902 & $3.06$ & $-2.44(11)$ & $-3.92(19)$ & & $-0.80(9)$ & $-4.5(2)$ & & $1.29(12)$ & $-4.47(19)$ & & $-2.4(2)$ & $-4.4(4)$ & (1) \\
J1012$-$4235 & $3.05$ & $-4.15(15)$ & $5.20(16)$ & & $-6.30(14)$ & $2.14(17)$ & & $-6.40(15)$ & $1.82(17)$ & & & & \\   
J1017$-$7156 & $3.15$ & $-7.43(8)$ & $6.91(7)$ & & $-8.93(6)$ & $-4.84(6)$ & & $-10.03(8)$ & $1.49(9)$ & & $-7.411(12)$ & $6.870(11)$ & (6) \\
J1024$-$0719 & $3.05$ & $-35.32(5)$ & $-48.39(12)$ & & $-14.37(4)$ & $-58.16(13)$ & & $8.62(7)$ & $-59.28(12)$ & & $-35.270(17)$ & $-48.22(3)$ & (6) \\
J1036$-$8317 & $2.86$ & $-11.35(6)$ & $2.86(6)$ & & $-1.83(6)$ & $-11.56(6)$ & & $-11.15(6)$ & $-3.57(5)$ & & & & \\   
J1045$-$4509 & $3.10$ & $-6.25(19)$ & $4.9(2)$ & & $-7.9(2)$ & $0.6(2)$ & & $-7.8(2)$ & $1.4(2)$ & & $-6.07(3)$ & $5.19(4)$ & (6) \\
J1101$-$6424 & $3.10$ & $-1.0(2)$ & $0.4(2)$ & & $-1.0(2)$ & $-0.5(3)$ & & $-1.1(3)$ & $0.0(3)$ & & & & \\   
J1103$-$5403 & $3.06$ & $-7.1(4)$ & $1.1(4)$ & & $-6.2(4)$ & $-3.8(4)$ & & $-7.1(4)$ & $-1.9(5)$ & & & & \\   
J1125$-$5825 & $3.15$ & $-8.95(13)$ & $1.79(13)$ & & $-7.81(13)$ & $-4.72(14)$ & & $-9.04(14)$ & $-1.21(15)$ & & $-10.0(3)$ & $2.4(3)$ & (7) \\
J1125$-$6014 & $3.06$ & $11.09(3)$ & $-13.00(3)$ & & $17.01(3)$ & $-1.57(4)$ & & $14.71(4)$ & $-8.69(4)$ & & $11.106(13)$ & $-13.037(14)$ & (6) \\
J1216$-$6410 & $3.15$ & $-7.80(5)$ & $2.69(5)$ & & $-7.54(5)$ & $-3.36(5)$ & & $-8.10(5)$ & $1.60(5)$ & & & & \\   
J1421$-$4409 & $3.10$ & $-10.27(13)$ & $-6.15(14)$ & & $-7.28(10)$ & $-9.50(15)$ & & $-11.78(15)$ & $-2.16(11)$ & & $-11.6(4)$ & $-7.9(8)$ & (8) \\
J1431$-$5740 & $3.17$ & $0.0(3)$ & $2.5(4)$ & & $-1.0(3)$ & $2.3(5)$ & & $1.0(5)$ & $2.3(4)$ & & & & \\   
J1435$-$6100 & $3.15$ & $-5.6(2)$ & $-2.4(3)$ & & $-4.00(18)$ & $-4.5(3)$ & & $-6.1(3)$ & $0.0(3)$ & & & & \\   
J1446$-$4701 & $3.17$ & $-4.24(9)$ & $-2.45(16)$ & & $-3.15(8)$ & $-3.73(18)$ & & $-4.88(14)$ & $-0.32(14)$ & & $-4.36(4)$ & $-3.00(7)$ & (6) \\
J1455$-$3330 & $3.06$ & $7.66(13)$ & $-2.3(2)$ & & $8.01(9)$ & $0.1(2)$ & & $5.5(2)$ & $-5.82(17)$ & & $7.98(8)$ & $-2.0(2)$ & (1) \\
J1525$-$5545 & $3.06$ & $-6.8(5)$ & $-3.3(8)$ & & $-5.5(5)$ & $-5.5(7)$ & & $-7.6(8)$ & $1.1(7)$ & & & & \\   
J1543$-$5149 & $3.15$ & $-4.33(18)$ & $-2.5(2)$ & & $-3.54(16)$ & $-3.5(3)$ & & $-4.9(3)$ & $0.68(18)$ & & $-4.3(14)$ & $-4(2)$ & (7) \\
J1545$-$4550 & $3.10$ & $-0.51(7)$ & $2.50(13)$ & & $-1.09(6)$ & $2.29(12)$ & & $1.12(12)$ & $2.23(11)$ & & $-0.48(2)$ & $2.37(4)$ & (6) \\
J1547$-$5709 & $2.99$ & $-6.59(18)$ & $-5.9(2)$ & & $-4.77(17)$ & $-7.4(3)$ & & $-8.8(3)$ & $-0.5(2)$ & & & & \\   
J1600$-$3053 & $3.12$ & $-1.04(6)$ & $-7.5(2)$ & & $0.48(4)$ & $-7.5(2)$ & & $-5.91(18)$ & $-4.79(12)$ & & $-0.960(7)$ & $-6.96(3)$ & (6) \\
J1603$-$7202 & $3.16$ & $-2.55(10)$ & $-7.40(11)$ & & $-0.21(8)$ & $-7.83(11)$ & & $-6.92(12)$ & $-3.66(9)$ & & $-2.447(11)$ & $-7.356(13)$ & (6) \\
J1614$-$2230 & $3.05$ & $4.4(5)$ & $-29(3)$ & & $9.52(7)$ & $-28(3)$ & & $-18(3)$ & $-23.1(17)$ & & $3.8(1)$ & $-32.5(7)$ & (1) \\
J1629$-$6902 & $3.16$ & $-6.50(3)$ & $-8.44(4)$ & & $-4.45(3)$ & $-9.68(4)$ & & $-10.62(4)$ & $-0.78(3)$ & & & & \\   
J1643$-$1224 & $3.05$ & $6.1(4)$ & $3.4(17)$ & & $5.6(3)$ & $4.0(19)$ & & $6^{+2}_{-3}$ & $-3.2^{+1.1}_{-1.3}$ & & $5.970(18)$ & $3.77(8)$ & (6) \\
J1652$-$4838 & $2.33$ & $-3.8(3)$ & $-9.7(7)$ & & $-2.5(3)$ & $-10.1(8)$ & & $-9.8(7)$ & $-3.2(5)$ & & & & \\   
J1658$-$5324 & $2.82$ & $0.02(10)$ & $2.86(17)$ & & $-0.33(11)$ & $2.84(17)$ & & $2.27(13)$ & $1.74(15)$ & & $0.02(10)$ & $4.90(23)$ & (9) \\
J1708$-$3506 & $2.67$ & $-5.8(3)$ & $-0.4(8)$ & & $-5.8(2)$ & $-0.9(8)$ & & $-3.8(7)$ & $4.5(5)$ & & $-5.3(8)$ & $-2(3)$ & (7) \\
J1713+0747 & $1.98$ & $4.94(4)$ & $-4.07(9)$ & & $5.30(4)$ & $-3.59(9)$ & & $-1.45(8)$ & $-6.24(7)$ & & $4.9254(14)$ & $-3.917(3)$ & (6) \\
J1719$-$1438 & $3.03$ & $3.76(10)$ & $-3.3(5)$ & & $3.99(9)$ & $-3.0(5)$ & & $-0.8(6)$ & $-5.0(3)$ & & $1.9(4)$ & $-11(2)$ & (7) \\
J1730$-$2304 & $3.03$ & $20.3(11)$ & $0(21)$ & & $20.29(6)$ & $1(21)$ & & $11(19)$ & $-17(11)$ & & $20.06(12)$ & $-4(2)$ & (6) \\ 
J1732$-$5049 & $3.22$ & $-0.60(9)$ & $-10.0(2)$ & & $-0.07(9)$ & $-10.0(2)$ & & $-8.8(2)$ & $-4.78(13)$ & & $-0.41(9)$ & $-9.87(19)$ & (2) \\
J1737$-$0811 & $3.03$ & $-3.9(3)$ & $-11.8(13)$ & & $-3.4(3)$ & $-11.4(10)$ & & $-12.5^{+1.5}_{-1.8}$ & $-2.6^{+0.8}_{-0.9}$ & & & & \\   
J1744$-$1134 & $3.03$ & $18.85(2)$ & $-9.73(13)$ & & $19.11(2)$ & $-9.21(13)$ & & $1.11(12)$ & $-21.18(7)$ & & $18.803(4)$ & $-9.390(18)$ & (6) \\
J1747$-$4036 & $3.14$ & $-1.3(2)$ & $0.0(7)$ & & $-1.3(2)$ & $-0.1(6)$ & & $-0.8(10)$ & $1.2(7)$ & & $-1.30(13)$ & $-2.7(4)$ & (15) \\
J1751$-$2857 & $3.03$ & $-7.37(10)$ & $-4.8(10)$ & & $-7.30(10)$ & $-5.0(10)$ & & $-7.8(9)$ & $3.9(5)$ & & $-7.4(1)$ & $-4.3(12)$ & (5) \\
J1757$-$5322 & $2.96$ & $-2.48(8)$ & $-10.02(16)$ & & $-2.42(8)$ & $-10.04(16)$ & & $-10.07(17)$ & $-2.33(9)$ & & & & \\   
J1801$-$1417 & $3.03$ & $-10.67(10)$ & $-1.8(6)$ & & $-10.68(10)$ & $-1.9(6)$ & & $-6.8^{+0.7}_{-0.6}$ & $8.4^{+0.4}_{-0.3}$ & & $-10.89(12)$ & $-3.0(10)$ & (5) \\
J1825$-$0319 & $3.06$ & $3.1(2)$ & $-1.6(6)$ & & $3.1(2)$ & $-1.8(6)$ & & $0.0(5)$ & $-3.5(4)$ & & & & \\   
J1832$-$0836 & $3.05$ & $-8.06(8)$ & $-21.5(4)$ & & $-9.30(8)$ & $-21.0(4)$ & & $-22.8(4)$ & $-2.8(2)$ & & $-8.06(5)$ & $-21.01(19)$ & (6) \\
J1843$-$1113 & $3.03$ & $-2.02(7)$ & $-2.9(4)$ & & $-2.23(7)$ & $-2.7(4)$ & & $-3.5(3)$ & $0.5(2)$ & & $-1.91(7)$ & $-3.2(3)$ & (5) \\
J1902$-$5105 & $3.08$ & $-3.90(11)$ & $-8.5(3)$ & & $-4.91(11)$ & $-8.0(3)$ & & $-9.2(3)$ & $1.43(14)$ & & $-4.8(13)$ & $4.4(16)$ & (9) \\
J1903$-$7051 & $3.17$ & $-6.56(13)$ & $-17.68(18)$ & & $-9.35(13)$ & $-16.37(18)$ & & $-18.31^{+0.20}_{-0.19}$ & $4.45^{+0.14}_{-0.13}$ & & $-8.8(16)$ & $-16(2)$ & (9) \\
J1909$-$3744 & $3.26$ & $-9.513(12)$ & $-35.70(4)$ & & $-13.852(11)$ & $-34.25(4)$ & & $-36.83(4)$ & $-3.06(2)$ & & $-9.5146(8)$ & $-35.776(3)$ & (6) \\
J1918$-$0642 & $3.06$ & $-7.02(5)$ & $-6.37(17)$ & & $-7.84(4)$ & $-5.33(17)$ & & $-8.83(14)$ & $3.44(11)$ & & $-7.15(2)$ & $-5.97(5)$ & (1) \\
J1933$-$6211 & $3.19$ & $-5.58(4)$ & $11.05(7)$ & & $-3.19(3)$ & $11.96(7)$ & & $10.58(7)$ & $6.43(4)$ & & $-5.54(7)$ & $10.7(2)$ & (10) \\
\hline
\end{tabular}
\end{threeparttable}
\label{tab:proper_motions}
\end{table*}

\begin{table*}
\centering
\contcaption{\textit{continued}}
\hspace*{-1.5cm}
\begin{threeparttable}
\begin{tabular}{@{}llllllllllllll}
\hline
\multicolumn{1}{c}{} & \multicolumn{1}{c}{} & \multicolumn{2}{c}{Equatorial coordinates} & \multicolumn{1}{c}{} & \multicolumn{2}{c}{Ecliptic coordinates} & \multicolumn{1}{c}{} & \multicolumn{2}{c}{Galactic coordinates} & \multicolumn{1}{c}{} & \multicolumn{3}{c}{Best Previous Measurement}\\
\cline{3-4}\cline{6-7}\cline{9-10}\cline{12-14}

Pulsar & Span & $\mu_\alpha\equiv\dot{\alpha}\cos\delta$ & $\mu_{\delta}$ & & $\mu_\lambda\equiv\dot{\lambda}\cos\beta$ & $\mu_{\beta}$ & & $\mu_{l}\equiv\dot{l}\cos b$ & $\mu_{b}$ & & $\mu_\alpha$ & $\mu_{\delta}$ & Ref \\
 & (yr) & (mas\,yr$^{-1}$) & (mas\,yr$^{-1}$) & & (mas\,yr$^{-1}$) & (mas\,yr$^{-1}$) & & (mas\,yr$^{-1}$) & (mas\,yr$^{-1}$) & & (mas\,yr$^{-1}$) & (mas\,yr$^{-1}$) &  \\
\hline
\multicolumn{14}{c}{Total Proper Motion Detections $>3\sigma$}\\
\hline
J1946$-$5403 & $3.01$ & $-1.08(3)$ & $-4.75(5)$ & & $-2.06(2)$ & $-4.41(5)$ & & $-4.84(5)$ & $0.55(3)$ & & & & \\   
J2010$-$1323 & $3.07$ & $2.63(8)$ & $-6.1(4)$ & & $1.25(3)$ & $-6.5(4)$ & & $-4.5(3)$ & $-4.9(2)$ & & $2.4(3)$ & $-5.6(3)$ & (11) \\
J2039$-$3616 & $3.01$ & $-10.11(5)$ & $-8.66(15)$ & & $-12.06(4)$ & $-5.64(16)$ & & $-10.17^{+0.15}_{-0.14}$ & $8.60(8)$ & & & & \\   
J2124$-$3358 & $3.07$ & $-14.17(6)$ & $-50.28(14)$ & & $-29.76(4)$ & $-42.93(15)$ & & $-51.44(15)$ & $9.11(8)$ & & $-14.109(19)$ & $-50.36(4)$ & (6) \\
J2129$-$5721 & $3.12$ & $9.30(4)$ & $-9.61(5)$ & & $4.54(3)$ & $-12.59(5)$ & & $-12.06(6)$ & $-5.76(3)$ & & $9.30(1)$ & $-9.576(13)$ & (6) \\
J2145$-$0750 & $3.07$ & $-9.7(3)$ & $-8.2(10)$ & & $-11.87(8)$ & $-4.5(10)$ & & $-11.9(7)$ & $4.4(8)$ & & $-9.48(2)$ & $-9.11(7)$ & (6) \\
J2150$-$0326 & $2.82$ & $6.5(2)$ & $-11.1(6)$ & & $2.28(9)$ & $-12.7(6)$ & & $-5.5(5)$ & $-11.3(7)$ & & & & \\
J2222$-$0137 & $3.07$ & $44.8(2)$ & $-5.7(6)$ & & $39.63(9)$ & $-21.7(7)$ & & $23.7(4)$ & $-38.5(6)$ & & $44.70(4)$ & $-5.69(8)$ & (12) \\
J2229$+$2643 & $2.54$ & $-1.70(9)$ & $-5.75(14)$ & & $-4.05(8)$ & $-4.42(15)$ & & $-4.70(9)$ & $-3.74(15)$ & & $-2.1(6)$ & $-5.7(5)$ & (1) \\
J2234$+$0944 & $2.54$ & $6.9(2)$ & $-33.2(5)$ & & $-6.48(13)$ & $-33.3(5)$ & & $-16.6(3)$ & $-29.6(5)$ & & $6.9(2)$ & $-32.0(4)$ & (1) \\
J2241$-$5236 & $3.19$ & $18.875(12)$ & $-5.286(13)$ & & $13.830(10)$ & $-13.888(15)$ & & $-14.904(17)$ & $-12.736(10)$ & & $18.881(4)$ & $-5.294(5)$ & (6) \\
J2317$+$1439 & $2.54$ & $-1.45(9)$ & $3.7(2)$ & & $0.20(6)$ & $4.0(2)$ & & $0.51(6)$ & $4.0(2)$ & & $-1.3(3)$ & $3.6(5)$ & (11) \\
J2322$+$2057 & $2.54$ & $-18.06(10)$ & $-15.24(15)$ & & $-22.83(7)$ & $-6.09(17)$ & & $-22.82(8)$ & $-6.14(17)$ & & $-18.4(4)$ & $-15.4(5)$ & (5) \\
J2322$-$2650 & $3.06$ & $-2.37(9)$ & $-8.20(17)$ & & $-5.59(6)$ & $-6.44(19)$ & & $-8.40(16)$ & $1.52(11)$ & & $-2.4(2)$ & $-8.3(4)$ & (13) \\
\hline
\multicolumn{14}{c}{Total Proper Motion Detections $<3\sigma$}\\
\hline
J0955$-$6150 & $3.12$ & $-0.08^{+0.15}_{-0.14}$ & $-0.21^{+0.16}_{-0.15}$ & & $0.12^{+0.16}_{-0.17}$ & $-0.19^{+0.15}_{-0.14}$ & & $0.06^{+0.17}_{-0.16}$ & $-0.21(15)$ & & & & \\
J1022$+$1001 & $2.97$ & $-27^{+50}_{-53}$ & $-27^{+129}_{-136}$ & & $-15.70(13)$ & $-36^{+140}_{-137}$ & & $9^{+84}_{-85}$ & $-38^{+111}_{-109}$ & & $-14.92^{+0.05}_{-0.03}$ & $-5.61(3)$ & (11) \\
J1327$-$0755 & $2.34$ & $-4(2)$ & $5(6)$ & & $-5.06^{+0.10}_{-0.11}$ & $3(6)$ & & $-2(4)$ & $6(5)$ & & & & \\
J1653$-$2054 & $3.04$ & $-10.5(13)$ & $-15(11)$ & & $-8.7^{+0.4}_{-0.3}$ & $-16^{+11}_{-12}$ & & $-19(10)$ & $-1(6)$ & & & & \\
J1705$-$1903 & $2.77$ & $-1.7(9)$ & $-1^{+7}_{-6}$ & & $-1.5(5)$ & $-1(7)$ & & $-2(6)$ & $1^{+4}_{-3}$ & & & & \\
J1721$-$2457 & $2.73$ & $3.4(6)$ & $5(8)$ & & $3.1(3)$ & $5(8)$ & & $6(7)$ & $0(4)$ & & & & \\
J1802$-$2124 & $3.16$ & $-1.6(3)$ & $-14^{+9}_{-8}$ & & $-1.7(3)$ & $-14(8)$ & & $-13(7)$ & $-6(4)$ & & $-1.13(12)$ & $-3(4)$ & (5) \\
J1811$-$2405 & $3.12$ & $0.53(9)$ & $3(5)$ & & $0.59(4)$ & $3(4)$ & & $3(4)$ & $1(2)$ & & $0.53(6)$ & $0$ & (14) \\
\hline
\end{tabular}
\begin{tablenotes}
\item References: 
(1) \cite{abb+18},
(2) \cite{rhc+16},
(3) \cite{vbc+22},
(4) \cite{gkr+19},
(5) \cite{dcl+16},
(6) \cite{rsc+21},
(7) \cite{nbb+14},
(8) \cite{sfj+20},
(9) \cite{ckr+15},
(10) \cite{gvo+17},
(11) \cite{dgb+19},
(12) \cite{gfg+21},
(13) \cite{sbb+18},
(14) \cite{ngf+20},
(15) \cite{aab+21a}
\end{tablenotes}
\end{threeparttable}
\end{table*}

\subsection{Parallaxes}\label{ch4:sec4:parallax}
\begin{figure}
\centering
 \includegraphics[width=0.5\textwidth]{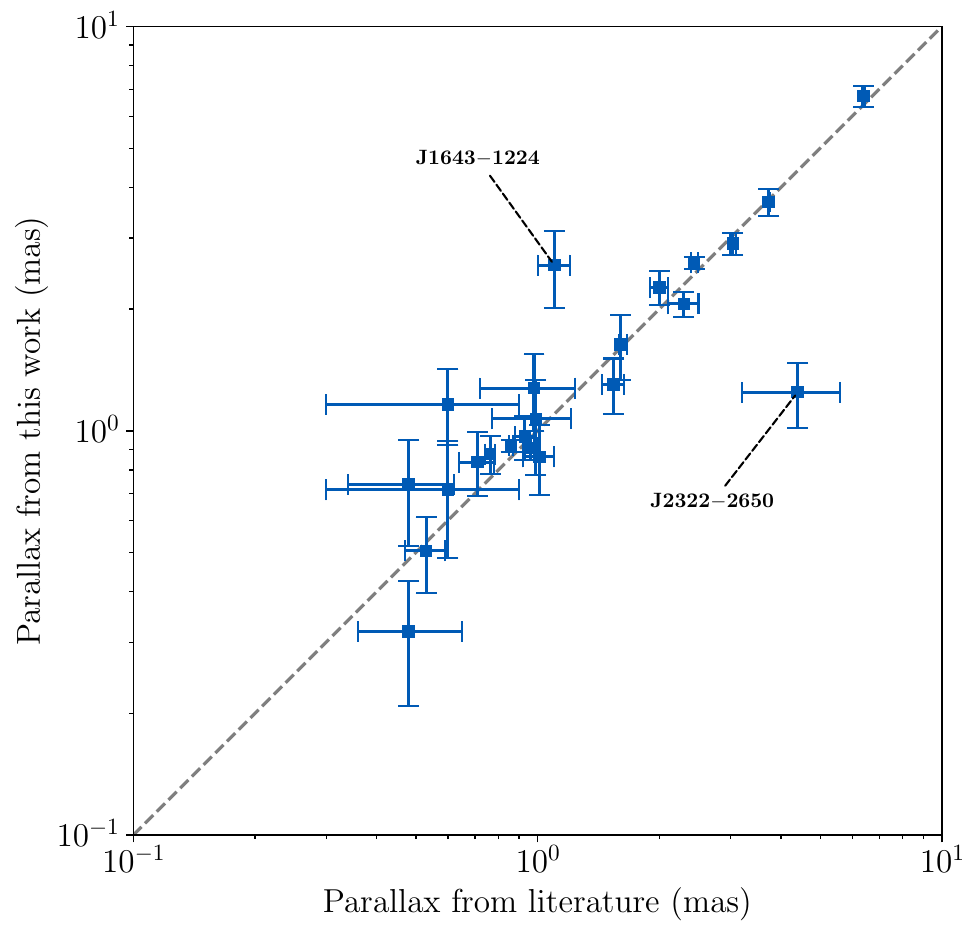}
\caption[Comparison of our parallax measurement with the best previous measurements]{Comparison of our $23$ parallax measurements with the measurements from literature. All of our measurements are consistent with previous work with only two exceptions, namely PSRs~J1643$-$1224 and J2322$-$2650 that we discuss in Section \ref{ch4:sec4:parallax}.}
\label{fig:parallax_log_log}
\end{figure}

The parallax measurements made in this study, as well as the best previous measurements of those pulsars in the literature and their corresponding measurement techniques (timing or VLBI), are listed in Table \ref{tab:parallax_distance} and shown in 
Figure \ref{fig:parallax_log_log} ($>3\sigma$ detections). The parallax values and their uncertainties are derived from the marginalized posterior probability distributions from \textsc{temponest}. There are $35$ of our MSPs that had parallax measurements with $>3\sigma$ significance. $12$ of these $35$ are new (PSRs~J0125$-$2327, J0614$-$3329, J0636$-$3044, J1421$-$4409, J1629$-$6902, J1652$-$4838, J1658$-$5324, J1757$-$5322, J1801$-$1417, J1811$-$2405, J1933$-$6211 and J1946$-$5403). We improved the precision of $4$ other parallaxes (of the $35$ MSPs) by a factor of 1.25 to 6 including PSRs~J1446$-$4701, J1455$-$3330, J2124$-$3358 and J2322$-$2650. For the MSPs with less than $3\sigma$ significance in parallax, we calculated the $68\%$ confidence upper level and listed them in the lower portion of Table \ref{tab:parallax_distance}. 

To gain confidence in our parallax measurements, we compared them with their values derived from other authors in Figure \ref{fig:parallax_log_log} as $22$ of our significant ($>3\sigma$) parallaxes have previous measurements. For PSR~J1713$+$0747, we measured the parallax of $0.88(10)$ mas using only $1.98$ yr of data as its subsequent timing was 
compromised by a sudden
step-change in its pulse profile and gradual decay back to the original profile. In addition to the previously measured parallax value of $0.763(21)$ mas by \citet{rsc+21} using timing analysis of the Parkes pulsar timing array second data release, a parallax of $0.82(3)$ mas was obtained by \citet{abb+18} using timing analysis of the NANOGrav 11-year data set. Using the VLBI technique, \citet{cbv+09} measured a value of $0.95^{+0.06}_{-0.05}$ mas. 
\cite{dcl+16} obtained a value of $0.90(3)$ mas, very close to our value.
All measurements are in agreement with our value to within about $1\sigma$.

Of the $35$ MSPs with significant measurements, only our parallaxes for PSRs~J1643$-$1224 and J2322$-$2650 (shown in Figure \ref{fig:parallax_log_log}) are markedly inconsistent with the best previous measurements. We discuss each of these two pulsars in turn below. 
Overall the standard deviation of the differences in relative parallaxes for the $23$ pulsars with previous measurements in literature is only $10$ percent, and if we exclude the two outliers discussed
above, only $7$ percent.

\subsubsection{PSR~J1643$-$1224}

For PSR~J1643$-$1224, we obtained the parallax value of $\varpi = 2.6^{+0.5}_{-0.6}$ mas which is quite inconsistent with the VLBI value of $\varpi = 1.1(1)$ mas obtained by \citet{dds+22} and also has the lower formal error of the measurements.

\citet{mnf+16} measured a value of $0.7(6)$ mas using the NANOGrav nine-year data set and a detailed model for DM variation that is known to be significant. They compared their value with the measurement of \citet{vbc+09}, who obtained a value of $2.2(4)$ mas which did not include a complex model for DM variation. \citet{mnf+16} hence argued that the limited DM variation modelling might be the possible source of the inconsistency between their measurements.

Two more measurements come from \citet{rsc+21}, who obtain the parallax to be $0.82(17)$ mas using the $24$ yr observations of the Parkes Pulsar Timing Array data set and \cite{dcl+16}, who obtain a parallax of $1.17(26)$ mas. Clearly, the timing of this pulsar presents challenges.

PSR~J1643$-$1224 has been shown to be located behind the H\textsc{II} region Sh 2$-$27 by \citet{hmg+11} and undergoes annual DM variations. Its timing with MeerKAT also exhibits strong evidence for chromatic variations (Miles et al. in prep), probably due to time-dependent scattering in the H\textsc{II} region
which are not modelled by our techniques.
\citet{occ20} reported that there is a significant excess DM for PSR~J1643$-$1224 consistent with its distance beyond the H\textsc{II} region.

We examined whether our parallax might be radio-frequency dependent, by measuring the parallax using just the top half of our band ($1284$--$1712$ MHz), and we obtained a smaller value of $1.7(4)$ mas, which is in better agreement ($1.5\sigma$) with the measurement of \citet{dds+22}. In our noise analysis for this pulsar, using the entire band, the noise model with the white, red, and DM noise parameters had the highest Bayesian evidence; however, using only the top part of the band the noise model with the white and red noise parameters had the highest Bayesian evidence. We know that this pulsar is significantly affected by interstellar scattering, but this is not currently included in our noise model. Weaknesses in our noise model such as this may lead to small biases in the parallax measurement. This would be good topic for a new study. For the rest of this paper, we adopt the VLBI parallax.

\subsubsection{PSR~J2322$-$2650}

For PSR~J2322$-$2650, we have measured a parallax of $1.3(2)$ mas, which is very different from the previously measured value of $4.4(12)$ mas by \citet{sbb+18} but since it has $6$ times lower uncertainty we favour our value. Our parallax distance for PSR~J2322$-$2650 is thus $0.80^{+0.18}_{-0.12}$ kpc which is 
much nearer the YMW16 distance of $0.76$ kpc than the $0.23$ kpc
suggested by \citet{sbb+18}.
We speculate that the inconsistency between our values is due to the absence of any DM noise model in the earlier analysis of \citet{sbb+18} and their poorer timing residuals.

\begin{table*}
\centering
\caption[Pulsar parallaxes and distances]{Pulsar parallaxes and distances. MSPs for which we have measured the parallax distances with statistical significance of $>3\sigma$ and $<3\sigma$ are listed in the upper and lower portions of the table, respectively. The DM distances, with two decimal digits, are derived from the two DM models: NE2001 and YMW16. $68\%$ upper limits on parallaxes and lower limits on parallax distances are provided for parallaxes with statistical significance of less than $3\sigma$ (lower portion).}
\hspace*{-1.5cm}
\begin{threeparttable}
\begin{tabular}{llllllllllll}
\hline
\multicolumn{1}{c}{} & \multicolumn{1}{c}{} & \multicolumn{2}{c}{} & \multicolumn{1}{c}{} & \multicolumn{3}{c}{Best Previous Measurement} & \multicolumn{1}{c}{} & \multicolumn{2}{c}{DM distances} & \multicolumn{1}{c}{}\\
\cline{6-8}\cline{10-11}
Pulsar & & Parallax $\varpi$ & Distance $d$ & & Parallax $\varpi$ & Technique & Ref. & &  NE2001 & YMW16 & \\
 & & (mas) & (kpc) & & (mas) & & & & (kpc) & (kpc)\\
\hline
\multicolumn{11}{c}{Significant Parallaxes Detections ($>3\sigma$)}\\
\hline
J0030$+$0451 & & $2.91(18)$ & $0.34(2)$ & & $3.04(5)$ & VLBI & (1) & & $0.32$ & $0.34$ \\
J0125$-$2327 & & $0.84(11)$ & $1.19^{+0.17}_{-0.13}$ & & & & & & $0.44$ & $0.87$ \\   
J0437$-$4715 & & $6.8(4)$ & $0.148^{+0.009}_{-0.008}$ & & $6.396(54)$ & VLBI & (2) & & $0.14$ & $0.16$ \\
J0613$-$0200 & & $0.86(17)$ & $1.16^{+0.28}_{-0.19}$ & & $1.01(9)$ & Timing & (3) & & $1.71$ & $1.02$ \\
J0614$-$3329 & & $1.5(4)$ & $0.67^{+0.25}_{-0.14}$ & & $<2.2$ & Timing & (4) & & $1.90$ & $2.69$ \\
J0636$-$3044 & & $4.3(7)$ & $0.23^{+0.04}_{-0.03}$ & & & & & & $1.00$ & $0.68$ \\   
J1024$-$0719 & & $0.97^{+0.12}_{-0.13}$ & $1.03^{+0.15}_{-0.11}$ & & $0.93(5)$ & VLBI & (1) & & $0.39$ & $0.38$ \\
J1125$-$6014 & & $1.2^{+0.3}_{-0.2}$ & $0.86^{+0.22}_{-0.16}$ & & $0.6(3)$ &Timing & (3) & & $1.49$ & $0.99$ \\ 
J1421$-$4409 & & $1.2(4)$ & $0.8^{+0.5}_{-0.2}$ & & & & & & $1.57$ & $2.09$ \\   
J1446$-$4701 & & $0.7(2)$ & $1.4^{+0.7}_{-0.3}$ & & $0.6(3)$ & Timing & (3) & & $1.46$ & $1.57$ \\
J1455$-$3330 & & $1.1(3)$ & $0.94^{+0.33}_{-0.19}$ & & $0.99(22)$ & Timing & (4) & & $0.53$ & $0.68$ \\
J1600$-$3053 & & $0.51(11)$ & $2.0^{+0.5}_{-0.3}$ & & $0.53(6)$ & Timing & (3) & & $1.63$ & $2.53$ \\
J1614$-$2230 & & $1.3(2)$ & $0.77^{+0.14}_{-0.11}$ & & $1.54(10)$ & Timing & (5) & & $1.27$ & $1.39$ \\
J1629$-$6902 & & $0.9(2)$ & $1.1^{+0.3}_{-0.2}$ & & & & & & $0.96$ & $0.96$ \\   
J1643$-$1224 & & $2.6^{+0.5}_{-0.6}$ & $0.39^{+0.11}_{-0.07}$ & & $1.1(1)$ & VLBI & (1) & & $2.40$ & $0.79$ \\
J1652$-$4838 & & $2.4^{+0.8}_{-0.7}$ & $0.41^{+0.19}_{-0.10}$ & & & & & & $3.38$ & $4.33$ \\   
J1658$-$5324 & & $1.3(3)$ & $0.75^{+0.17}_{-0.12}$ & & & & & & $0.93$ & $0.88$ \\   
J1713$+$0747 & & $0.88(10)$ & $1.14^{+0.15}_{-0.11}$ & & $0.763(21)$ & Timing & (3) & & $0.89$ & $0.92$ \\
J1730$-$2304 & & $2.3(2)$ & $0.44^{+0.05}_{-0.04}$ & & $2.0(1)$ & VLBI & (1) & & $0.53$ & $0.51$ \\
J1744$-$1134 & & $2.61(9)$ & $0.384^{+0.014}_{-0.013}$ & & $2.44(5)$ & Timing & (3) & & $0.41$ & $0.15$ \\
J1757$-$5322 & & $1.2(3)$ & $0.81^{+0.25}_{-0.15}$ & & & & & & $0.96$ & $0.94$ \\   
J1801$-$1417 & & $0.6(2)$ & $1.7^{+1.1}_{-0.5}$ & & &  & & & $1.52$ & $1.10$ \\         
J1811$-$2405 & & $0.72(15)$ & $1.4^{+0.4}_{-0.2}$ & & $<0.4$ & Timing & (4) & & $1.77$ & $1.83$ \\
J1832$-$0836 & & $0.7(2)$ & $1.4^{+0.6}_{-0.3}$ & & $0.48(14)$ & Timing & (5) & & $1.11$ & $0.81$ \\
J1909$-$3744 & & $0.92(3)$ & $1.09(4)$ & & $0.86(1)$ & Timing & (3) & & $0.46$ & $0.56$ \\
J1918$-$0642 & & $0.84(15)$ & $1.20^{+0.26}_{-0.18}$ & & $0.71(7)$ & VLBI & (1) & & $1.24$ & $1.03$ \\
J1933$-$6211 & & $0.7(2)$ & $1.5^{+0.6}_{-0.3}$ & & & & & & $0.52$ & $0.65$ \\   
J1946$-$5403 & & $0.81^{+0.10}_{-0.11}$ & $1.23^{+0.18}_{-0.14}$ & & & & & & $0.87$ & $1.15$ \\   
J2010$-$1323 & & $0.32(11)$ & $3.1^{+1.7}_{-0.8}$ & & $0.41(12)$ & Timing & (10) & & $1.02$ & $1.16$ \\
J2124$-$3358 & & $2.06^{+0.16}_{-0.15}$ & $0.48(4)$ & & $2.3(2)$ & Timing & (3) & & $0.27$ & $0.36$ \\
J2145$-$0750 & & $1.6(3)$ & $0.61^{+0.14}_{-0.09}$ & & $1.603^{+0.063}_{-0.009}$ & VLBI & (9) & & $0.57$ & $0.69$ \\
J2222$-$0137 & & $3.7(3)$ & $0.27(2)$ & & $3.730(16)$ & VLBI & (6) & & $0.31$ & $0.27$ \\
J2241$-$5236 & & $0.91(6)$ & $1.10^{+0.07}_{-0.06}$ & & $0.96(4)$ & Timing & (3) & & $0.51$ & $0.96$ \\
J2322$+$2057 & & $1.2(3)$ & $0.80^{+0.21}_{-0.15}$ & & $0.98(26)$ & Timing & (5) & & $0.80$ & $1.01$ \\
J2322$-$2650 & & $1.3(2)$ & $0.80^{+0.18}_{-0.12}$ & & $4.4(12)$ & Timing & (7) & & $0.32$ & $0.76$ \\
\hline
\end{tabular}
\end{threeparttable}
\label{tab:parallax_distance}
\end{table*}

\begin{table*}
\centering
\contcaption{}
\hspace*{-2cm}
\begin{threeparttable}
\begin{tabular}{llllllllllll}
\hline
\multicolumn{1}{c}{} & \multicolumn{2}{c}{Parallax $\varpi$} & \multicolumn{1}{c}{Distance $d$} & \multicolumn{1}{c}{} & \multicolumn{3}{c}{Best Previous Measurement} & \multicolumn{1}{c}{} & \multicolumn{2}{c}{DM distances} & \multicolumn{1}{c}{}\\
\cline{2-3}\cline{6-8}\cline{10-11}
Pulsar & Measurement & $1\sigma$ limit & $1\sigma$ limit & & Parallax $\varpi$ & Technique & Ref. & &  NE2001 & YMW16 & \\
 & (mas) & (mas) & (kpc) & & (mas) & & & & (kpc) & (kpc)\\
\hline
\multicolumn{11}{c}{Weak Parallaxes Detections ($<3\sigma$)}\\
\hline
J0610$-$2100 & $0.5^{+0.5}_{-0.4}$ & $<1.0$ & $>1.0$ & & $0.72(11)$ & VLBI & (1) & & $3.54$ & $3.26$ \\  
J0711$-$6830 & $4^{+6}_{-3}$ & $<10.0$ & $>0.1$ & & & & & & $0.86$ & $0.11$ \\     
J0900$-$3144 & $1.1^{+1.3}_{-0.8}$ & $<2.4$ & $>0.4$ & & $0.77(44)$ & Timing & (8) & & $0.54$ & $0.38$ \\  
J0931$-$1902 & $0.5^{+0.4}_{-0.3}$ & $<0.9$ & $>1.2$ & & $0.7(7)$ & Timing & (5) & & $1.88$ & $3.72$ \\  
J0955$-$6150 & $0.6^{+0.8}_{-0.4}$ & $<1.4$ & $>0.6$ & & & & & & $4.04$ & $2.17$ \\
J1012$-$4235 & $1.3^{+0.9}_{-0.8}$ & $<2.2$ & $>0.5$ & & & & & & $2.51$ & $0.37$ \\     
J1017$-$7156 & $1.5^{+0.9}_{-0.8}$ & $<2.4$ & $>0.5$ & & $0.6(6)$ & Timing & (3) & & $2.98$ & $1.81$ \\  
J1022$+$1001 & $1.1(4)$ & $<1.5$ & $>0.7$ & & $1.38^{+0.04}_{-0.03}$ & VLBI & (9) & & $0.45$ & $0.83$ \\
J1036$-$8317 & $0.6^{+0.7}_{-0.4}$ & $<1.3$ & $>0.7$ & & & & & & $1.03$ & $0.93$ \\     
J1045$-$4509 & $1.0^{+1.0}_{-0.7}$ & $<2.0$ & $>0.5$ & & $1.7(7)$ & Timing & (3) & & $1.96$ & $0.33$ \\  
J1101$-$6424 & $1.0^{+1.3}_{-0.7}$ & $<2.3$ & $>0.4$ & & & & & & $4.48$ & $2.18$ \\     
J1103$-$5403 & $1.4^{+1.9}_{-1.0}$ & $<3.3$ & $>0.3$ & & & & & & $2.55$ & $1.68$ \\     
J1125$-$5825 & $1.3^{+0.8}_{-0.7}$ & $<2.1$ & $>0.5$ & & & & & & $2.62$ & $1.74$ \\     
J1216$-$6410 & $0.9(4)$ & $<1.3$ & $>0.8$ & & $0.6(3)$ & Timing & (3) & & $1.33$ & $1.10$ \\  
J1327$-$0755 & $0.13^{+0.17}_{-0.09}$ & $<0.3$ & $>3.4$ & & & & & & $1.73$ & $25.00$ \\
J1431$-$5740 & $2.4^{+1.4}_{-1.3}$ & $<3.8$ & $>0.3$ & & & & & & $2.55$ & $3.55$ \\     
J1435$-$6100 & $0.4^{+0.5}_{-0.3}$ & $<0.9$ & $>1.1$ & & & & & & $2.16$ & $2.82$ \\     
J1525$-$5545 & $1.5^{+1.3}_{-1.0}$ & $<2.8$ & $>0.3$ & & & & & & $2.37$ & $3.14$ \\     
J1543$-$5149 & $0.6^{+0.5}_{-0.4}$ & $<1.1$ & $>0.9$ & & & & & & $2.43$ & $1.15$ \\     
J1545$-$4550 & $0.38(18)$ & $<0.6$ & $>1.8$ & & $0.45(14)$ & Timing & (3) & & $2.13$ & $2.25$ \\  
J1547$-$5709 & $0.6^{+0.6}_{-0.4}$ & $<1.2$ & $>0.8$ & & & & & & $1.87$ & $2.70$ \\     
J1603$-$7202 & $1.1^{+0.7}_{-0.6}$ & $<1.8$ & $>0.6$ & & $0.3(3)$ & Timing & (3) & & $1.17$ & $1.13$ \\  
J1653$-$2054 & $0.24^{+0.34}_{-0.17}$ & $<0.6$ & $>1.7$ & & & & & & $1.67$ & $2.63$ \\
J1705$-$1903 & $0.6^{+0.6}_{-0.4}$ & $<1.2$ & $>0.8$ & & & & & & $1.65$ & $2.34$ \\
J1708$-$3506 & $0.5^{+0.4}_{-0.3}$ & $<0.9$ & $>1.1$ & & & & & & $2.80$ & $3.32$ \\
J1719$-$1438 & $0.30^{+0.23}_{-0.18}$ & $<0.6$ & $>1.9$ & & & & & & $1.21$ & $0.34$ \\     
J1721$-$2457 & $0.24^{+0.29}_{-0.16}$ & $<0.6$ & $>2.0$ & & & & & & $1.30$ & $1.39$ \\
J1732$-$5049 & $0.7(4)$ & $<1.1$ & $>0.9$ & & & & & & $1.41$ & $1.87$ \\     
J1737$-$0811 & $0.4^{+0.4}_{-0.3}$ & $<0.8$ & $>1.2$ & & & & & & $1.71$ & $0.21$ \\  
J1747$-$4036 & $0.17^{+0.25}_{-0.12}$ & $<0.5$ & $>2.3$ & & $0.1(7)$ & Timing & (5) & & $3.39$ & $7.15$ \\  
J1751$-$2857 & $0.6(3)$ & $<0.9$ & $>1.2$ & & & & & & $1.10$ & $1.09$ \\    
J1802$-$2124 & $0.4^{+0.5}_{-0.3}$ & $<0.9$ & $>1.2$ & & $1.24(57)$ & Timing & (8) & & $2.94$ & $3.03$ \\ 
J1825$-$0319 & $0.5^{+0.4}_{-0.3}$ & $<0.9$ & $>1.1$ & & & & & & $3.07$ & $3.86$ \\     
J1843$-$1113 & $0.32^{+0.17}_{-0.16}$ & $<0.5$ & $>2.0$ & & $0.69(33)$ & Timing & (8) & & $1.70$ & $1.71$ \\  
J1902$-$5105 & $0.20^{+0.25}_{-0.14}$ & $<0.5$ & $>2.3$ & & & & & & $1.18$ & $1.65$ \\     
J1903$-$7051 & $0.8^{+0.8}_{-0.5}$ & $<1.6$ & $>0.6$ & & & & & & $0.76$ & $0.93$ \\     
J2039$-$3616 & $0.25^{+0.14}_{-0.13}$ & $<0.4$ & $>2.5$ & & & & & & $0.91$ & $1.70$ \\     
J2129$-$5721 & $0.12^{+0.13}_{-0.08}$ & $<0.3$ & $>4.1$ & & $0.26(17)$ & Timing & (3) & & $1.36$ & $6.16$ \\  
J2150$-$0326 & $0.30^{+0.28}_{-0.19}$ & $<0.6$ & $>1.7$ & & & & & & $1.05$ & $1.98$ \\     
J2229$+$2643 & $0.6(3)$ & $<0.9$ & $>1.1$ & & $0.2(3)$ & Timing & (5) & & $1.43$ & $1.80$ \\  
J2234$+$0944 & $1.0(4)$ & $<1.4$ & $>0.7$ & & $0.7(3)$ & Timing & (5) & & $1.00$ & $1.58$ \\
J2317$+$1439 & $0.29^{+0.16}_{-0.15}$ & $<0.5$ & $>2.2$ & & $0.60(8)$ & Timing & (5) & & $0.83$ & $2.16$ \\
\hline
\end{tabular}
\begin{tablenotes}
\item References: 
(1) \cite{dds+22},
(2) \cite{dvt+08},
(3) \cite{rsc+21},
(4) \cite{gsl+16},
(5) \cite{aab+21a},
(6) \cite{gfg+21},
(7) \cite{sbb+18},
(8) \cite{dcl+16},
(9) \cite{dgb+19},
(10) \cite{aab+21b}
\end{tablenotes}
\end{threeparttable}
\end{table*}

\begin{table}
\centering
\caption[Pulsar velocities]{Pulsar transverse velocities. The second column shows the velocities that are corrected for the peculiar velocity of the Sun and the Galactic differential rotation ($v_{\rm \perp,LSR}$). The third column shows the observed transverse velocities without any corrections ($v_{\rm \perp,obs}$). The fourth column shows the best previous reported velocities,
some of which did not correct for the local standard of rest. 
The size of the error bars is showing $68$\% confidence interval from posterior probability distributions of velocities.}
\begin{threeparttable}
\begin{tabular}{llllllc}
\hline
\multicolumn{1}{c}{} & \multicolumn{1}{c}{}& \multicolumn{1}{c}{}& \multicolumn{1}{c}{} & \multicolumn{1}{c}{} & \multicolumn{2}{c}{Best Previous Reports} \\
\cline{6-7}
Pulsar & $v_{\rm \perp,LSR}$ & & $v_{\rm \perp,obs}$  & & $v_{\perp}$ & Ref. \\
 & (km s$^{-1}$) & & (km s$^{-1}$) & & (km s$^{-1}$) & \\
\hline
J0030$+$0451 & $25^{+4}_{-3}$ & & $11^{+3}_{-2}$ & & $15.4(2)$ & (1) \\
J0125$-$2327 & $204^{+31}_{-24}$ & & $218^{+31}_{-24}$ & &  &  \\
J0437$-$4715 & $99(6)$ & & $99(6)$ & & $104.72(14)$ & (2) \\
J0610$-$2100$^{\dag}$ & $113^{+23}_{-17}$ & & $126^{+23}_{-17}$ & & $120^{+25}_{-17}$ & (1) \\
J0613$-$0200 & $58^{+14}_{-9}$ & & $58^{+14}_{-9}$ & & $49(5)$ & (3) \\
J0614$-$3329 & $22.4^{+2.3}_{-1.3}$ & & $6.1^{+2.3}_{-1.3}$ & &  &  \\
J0636$-$3044 & $43^{+6}_{-5}$ & & $36^{+6}_{-5}$ & &  &  \\
J1024$-$0719 & $278^{+44}_{-32}$ & & $293^{+44}_{-33}$ & & $300(20)$ & (1) \\
J1045$-$4509$^{\dag}$ & $16^{+4}_{-2}$ & & $22^{+15}_{-6}$ & & $22^{+16}_{-7}$ & (3) \\
J1125$-$6014 & $101^{+17}_{-12}$ & & $70^{+18}_{-13}$ & & $124^{+90}_{-40}$ & (3) \\
J1421$-$4409 & $21^{+27}_{-13}$ & & $48^{+27}_{-13}$ & &  &  \\
J1446$-$4701 & $7^{+8}_{-5}$ & & $33^{+15}_{-8}$ & & $39^{+40}_{-14}$ & (3) \\
J1455$-$3330 & $53^{+12}_{-7}$ & & $36^{+13}_{-7}$ & &  &  \\
J1545$-$4550$^{\dag}$ & $56^{+10}_{-5}$ & & $27^{+12}_{-7}$ & & $26^{+12}_{-7}$ & (3) \\
J1600$-$3053 & $57^{+20}_{-12}$ & & $71^{+20}_{-13}$ & & $62^{+8}_{-7}$ & (3) \\
J1614$-$2230 & $98^{+21}_{-16}$ & & $108^{+22}_{-17}$ & &  &  \\
J1629$-$6902 & $26^{+16}_{-10}$ & & $56^{+17}_{-11}$ & &  &  \\
J1643$-$1224$^{\dag}$ & $39^{+6}_{-5}$ & & $30^{+5}_{-4}$ & & $41^{+5}_{-4}$ & (1) \\
J1652$-$4838 & $5^{+7}_{-2}$ & & $20^{+9}_{-5}$ & & & \\
J1658$-$5324 & $31.3^{+2.7}_{-1.8}$ & & $10.1^{+2.7}_{-1.8}$ & &  &  \\
J1713$+$0747 & $28^{+4}_{-3}$ & & $35^{+4}_{-3}$ & & $39.1(11)$ & (3) \\
J1730$-$2304 & $54^{+26}_{-11}$ & & $53^{+25}_{-10}$ & & $54^{+3}_{-2}$ & (1) \\
J1744$-$1134 & $35.0(13)$ & & $38.6(14)$ & & $40.9^{+0.9}_{-0.8}$ & (3) \\
J1757$-$5322 & $21^{+12}_{-7}$ & & $39^{+12}_{-7}$ & &  &  \\
J1801$-$1417 & $84^{+57}_{-24}$ & & $85^{+57}_{-24}$ & &  &  \\
J1832$-$0836 & $134^{+61}_{-33}$ & & $149^{+61}_{-33}$ & & $170^{+80}_{-40}$ & (3) \\
J1843$-$1113$^{\dag}$ & $14^{+18}_{-5}$ & & $24^{+20}_{-8}$ & &  &  \\
J1909$-$3744 & $179(7)$ & & $191(7)$ & & $203(3)$ & (3) \\
J1918$-$0642 & $44^{+11}_{-8}$ & & $54^{+11}_{-8}$ & & $44^{+6}_{-5}$ & (1) \\
J1933$-$6211 & $117^{+37}_{-20}$ & & $88^{+37}_{-20}$ & &  &  \\
J1946$-$5403 & $21^{+3}_{-2}$ & & $28^{+4}_{-3}$ & &  &  \\
J2010$-$1323 & $83^{+47}_{-22}$ & & $99^{+49}_{-25}$ & & $58.7^{+27}_{-16}$ & (4) \\
J2124$-$3358 & $114^{+10}_{-9}$ & & $120^{+10}_{-9}$ & & $110^{+11}_{-9}$ & (3) \\
J2145$-$0750 & $37^{+8}_{-5}$ & & $37^{+8}_{-6}$ & & $38.9^{+0.5}_{-1.9}$ & (4) \\
J2222$-$0137 & $47^{+5}_{-4}$ & & $58^{+5}_{-4}$ & &  &  \\
J2234$+$0944$^{\dag}$ & $207^{+162}_{-68}$ & & $227^{+162}_{-69}$ & &  &  \\
J2241$-$5236 & $79^{+7}_{-6}$ & & $103^{+7}_{-6}$ & & $97(5)$ & (3) \\
J2322$+$2057 & $82^{+23}_{-17}$ & & $90^{+23}_{-17}$ & &  &  \\
J2322$-$2650 & $32^{+7}_{-4}$ & & $32^{+8}_{-5}$ & &  &  \\
\hline
\end{tabular}
\begin{tablenotes}
\item[$\bullet$] We used our proper motions and other author's 
(usually parallax) distances (listed in Table \ref{tab:parallax_distance}) to calculate the velocities of the MSPs that are marked by a ``\dag''.
\item[$\bullet$] References: 
(1) \cite{dds+22},
(2) \cite{rhc+16},
(3) \cite{rsc+21},
(4) \cite{dgb+19},
\end{tablenotes}
\end{threeparttable}
\label{tab:velocities}
\end{table}

\section{Kinematics of the MSP sample}\label{sec:results-analysis}

In this section we study the kinematics of our MSP sample, using only those with parallax-based distances and velocities. While models of the ISM can be used to estimate distances to pulsars, it is well known that this can lead to large uncertainties. We choose here to base our kinematic analysis on MSPs with high quality distances ($>3\sigma$ significance) only. 

For a detailed analysis of the uncertainties involved in the use of DM-based distances via ISM models, we refer the reader to Appendix \ref{sec:dist_comparison_dist_ratio}.

\subsection{Distances}
We used the posterior samples of the parallax measurements to derive the posterior distribution of the distances. The distances of the $35$ MSPs with $>3\sigma$ significance are listed in the upper portion of Table \ref{tab:parallax_distance}. For the MSPs with $<3\sigma$ significance in parallax, we calculated $68\%$ confidence lower limit on distances and listed them in the lower portion of Table \ref{tab:parallax_distance}. 

\citet{lk73} pointed out a bias in the parallax of stars and showed that the bias depends on the significance of the parallax with low-significance parallaxes more likely to be over-estimated due to volume effects.
This has the effect of pushing low-significance pulsar parallax objects further from the Sun.
\citet{vlm10,vwc+12} confirmed the existence of the Lutz-Kelker (LK) bias in observations, and they provided a method in order to correct the values of parallaxes and distances. For our discussions concerning pulsar
distances and velocities, we will not adopt the LK corrections for individual objects to ease comparison with
other recent authors \cite[e.g., ][]{mnf+16,abb+18}, but will discuss what effects it would have on the median velocities of the population. 
One problem with the LK correction is that the magnitude of the correction depends upon the assumed model for the spatial (i.e.\ Galactic) distribution of the pulsars, and 
also the luminosity function of MSPs. For high-precision parallaxes, the effects are negligible.

\subsection{Tangential Velocities}

From the proper motions and distances of the MSPs, we calculated the barycentric transverse velocities, $v_{\perp}$, and then using the method explained in Section \ref{sec:method_vel_corr} we corrected them for the peculiar velocity of the Sun and the rotation curve of the Galaxy, assuming zero radial velocity with respect to our line of sight. In Table \ref{tab:velocities}, a summary of $39$ velocities for two subsets of our MSPs is given. The first subset is those MSPs with $>3\sigma$ significance for their parallax and proper motion measurements. The second subset is those MSPs with $>3\sigma$ significance for their proper motions measurements but $<3\sigma$ significance for their parallax measurements. For MSPs in the second subset namely: PSRs~J0610$-$2100, J1045$-$4509, J1545$-$4550, J1643$-$1224, J1843$-$1113, and J2234$+$0944, we used parallaxes measured by other authors listed in the fifth column of Table \ref{tab:parallax_distance} to calculate their transverse velocities (these velocities are denoted by a $\dag$ after the pulsar name). We used the positions, proper motions, and parallaxes of other MSPs in the pulsar catalogue \citep{mht+05} and selected those with significant parallaxes ($>3\sigma$) and calculated their transverse velocities using our method described in Section \ref{sec:methods} in order to increase our MSP velocity sample.

\subsubsection{Isolated versus Binary Millisecond Pulsars}
\begin{figure}
\centering
 \includegraphics[width=0.5\textwidth]{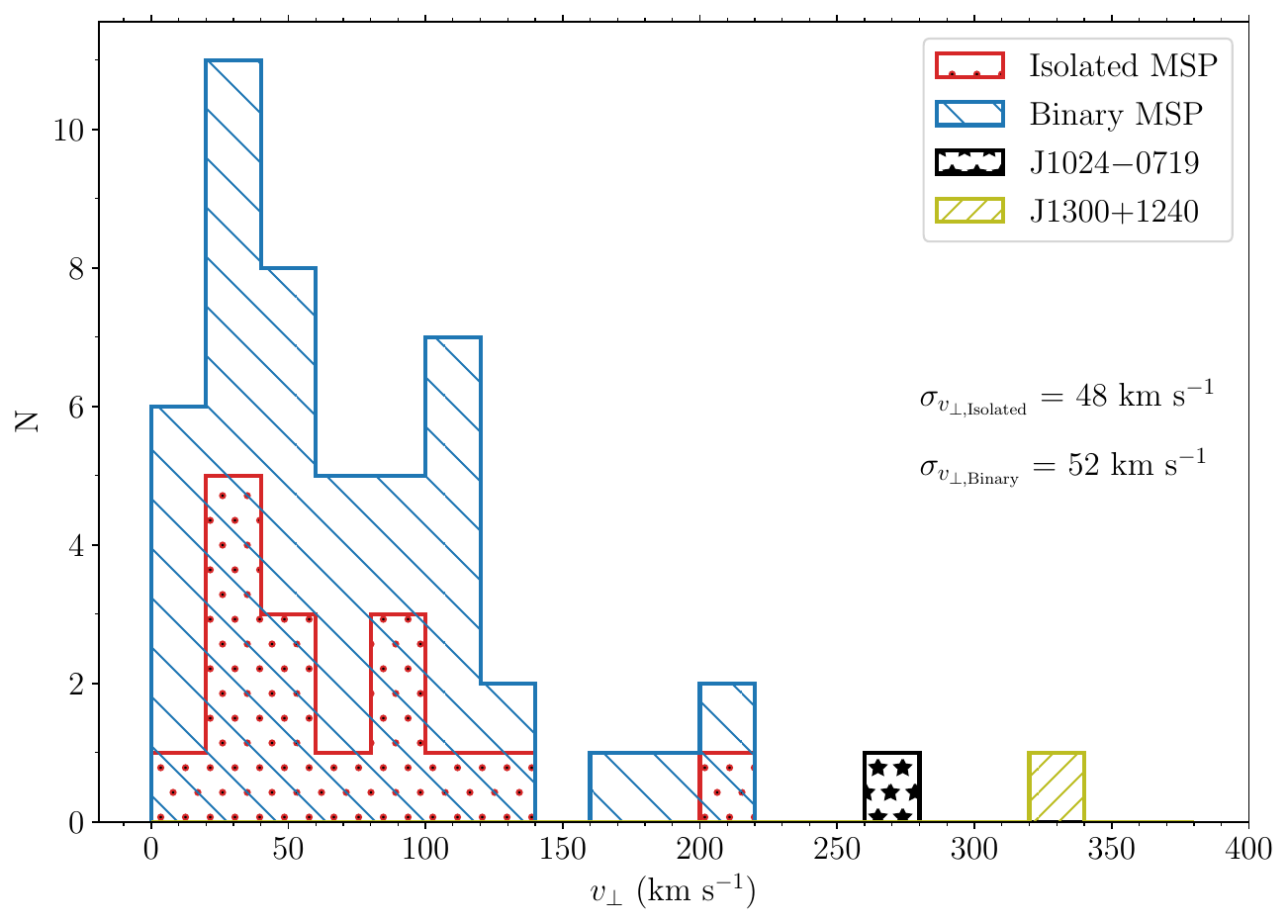}
\caption[Histograms of transverse velocities for the isolated and binary MSPs]{Histograms of transverse velocities for the isolated MSPs (red) and the binary MSPs (blue). PSR~J1024$-$0719 is shown in black due to its extremely long orbital period ($2$--$20$ kyr \citep{kkn+16}). PSR~J1300$+$1240 is shown in olive as it is a planetary system \citep{wol90a}. These two pulsars are not included in obtaining the velocity dispersion of binary MSPs as the evolutionary history is quite unlike that of the others in the sample.}
\label{fig:vel_hist_isolated_binary}
\end{figure}

We divided the MSPs into two subsets depending upon whether they are isolated or binary MSPs. The velocity distributions for the isolated and binary MSPs are presented in Figure \ref{fig:vel_hist_isolated_binary}. We obtained a mean velocity of $67(12) \, {\rm km \, s^{-1}}$ for $16$ isolated MSPs, and a mean velocity of $72(8) \, {\rm km \, s^{-1}}$ for the $49$ binary MSPs. 
Between these MSPs, PSR~J1024$-$0719 is a special case with an extremely wide binary with the orbital period of $2$--$20$ kyr \citep{kkn+16} (presented with a black bin in Figure  \ref{fig:vel_hist_isolated_binary}). Also, PSR~J1300$+$1240 is a planetary system \citep{wol90a} (presented with an olive bin in Figure  \ref{fig:vel_hist_isolated_binary}). We did not include them for calculating the mean velocity of the binary MSPs as their evolutionary history is probably quite different from the other systems. The two-sample Kolmogorov-Smirnov test was performed for comparing the distributions of velocity samples. The maximum absolute difference between the cumulative distribution functions of the two samples was $0.17$ and the $p$-value was $0.87$. This means with our limited sample that there is no statistical evidence that the two samples are drawn from different distributions. Whatever causes isolated MSPs to be created does
not have an appreciably different effect on their velocities
from that of the binaries.

\subsection{Dispersion of the Velocity Components}\label{sec:velocity_dispersion}
\begin{figure}
\centering
 \includegraphics[width=0.5\textwidth]{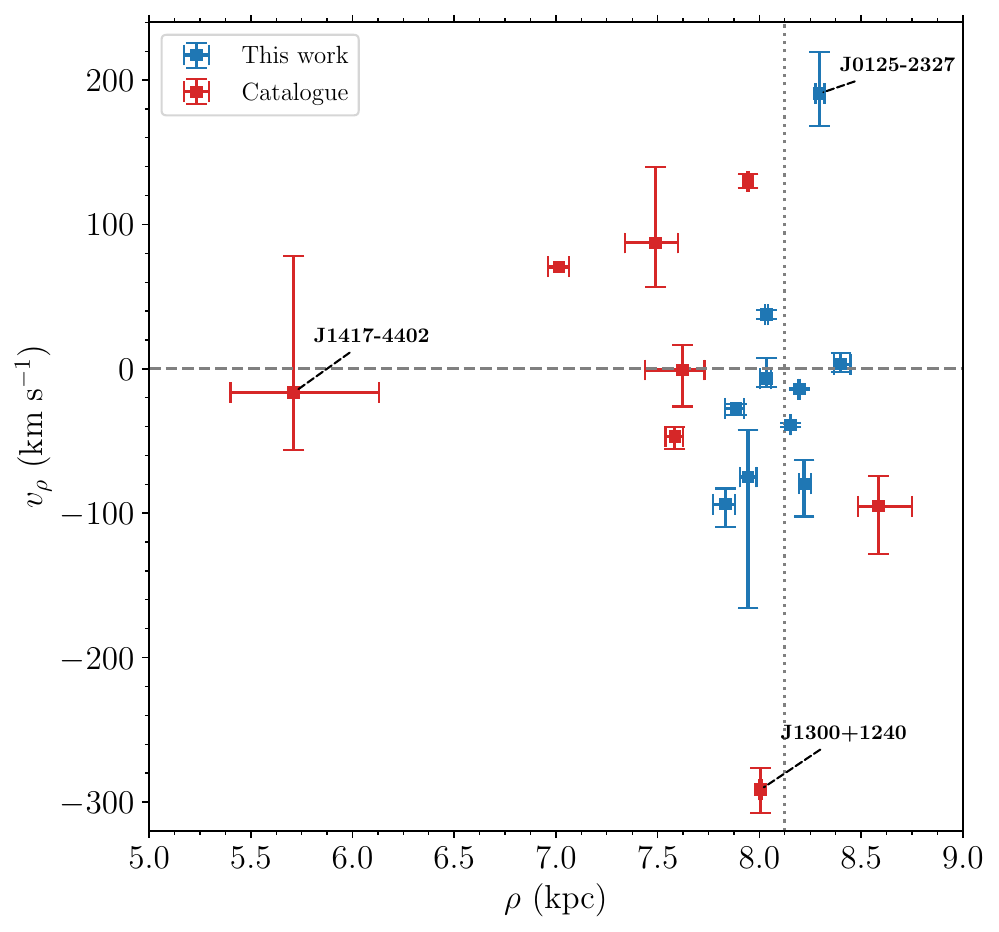}
\caption[Radial component of the cylindrical Galactocentric velocity versus distance from the Galactic centre]{Radial component of the cylindrical Galactocentric velocity ($v_{\rho}$) versus distance from the Galactic centre ($\rho$) for MSPs whose radial velocity vectors are $> 70^{\circ}$ from the line of sight direction. The dotted line indicates the assumed position of the Sun. Velocities calculated using MSPs in this work are in blue and from the pulsar catalogue are in red.  The velocity dispersion of these MSPs is $63(11)$ km s$^{-1}$, if we exclude PSR~J0125$-$2327 and J1300$+$1240 as outliers.}
\label{fig:velocity_rho}
\end{figure}

\begin{figure}
\centering
 \includegraphics[width=0.5\textwidth]{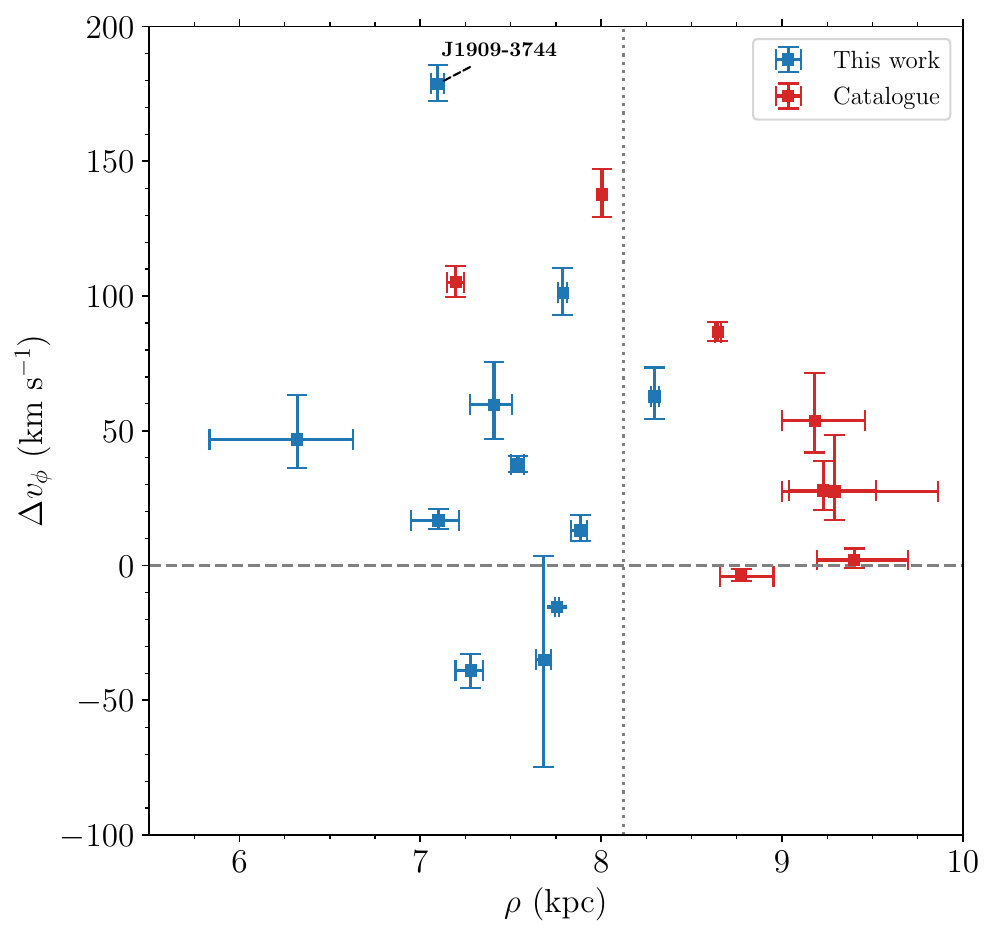}
\caption[Rotational component of the cylindrical Galactocentric velocity versus distance from the galactic centre]{Rotational component of the cylindrical Galactocentric velocity ($\Delta v_{\phi}$) versus distance from the Galactic centre ($\rho$) for MSPs whose rotational velocity vector are $> 70^{\circ}$ from the line of sight direction. The dotted line indicates the assumed position of the Sun. Velocities calculated using MSPs in this work are in blue and from the pulsar catalogue are in red. The velocity dispersion of these MSPs is this component is $46(8)$ km s$^{-1}$, excluding PSR~J1909$-$3744 as an outlier. The signature of asymmetric drift is visible with the mean drift velocity of $39(11)$ km s$^{-1}$.}
\label{fig:velocity_phi}
\end{figure}

\begin{figure}
\centering
 \includegraphics[width=0.5\textwidth]{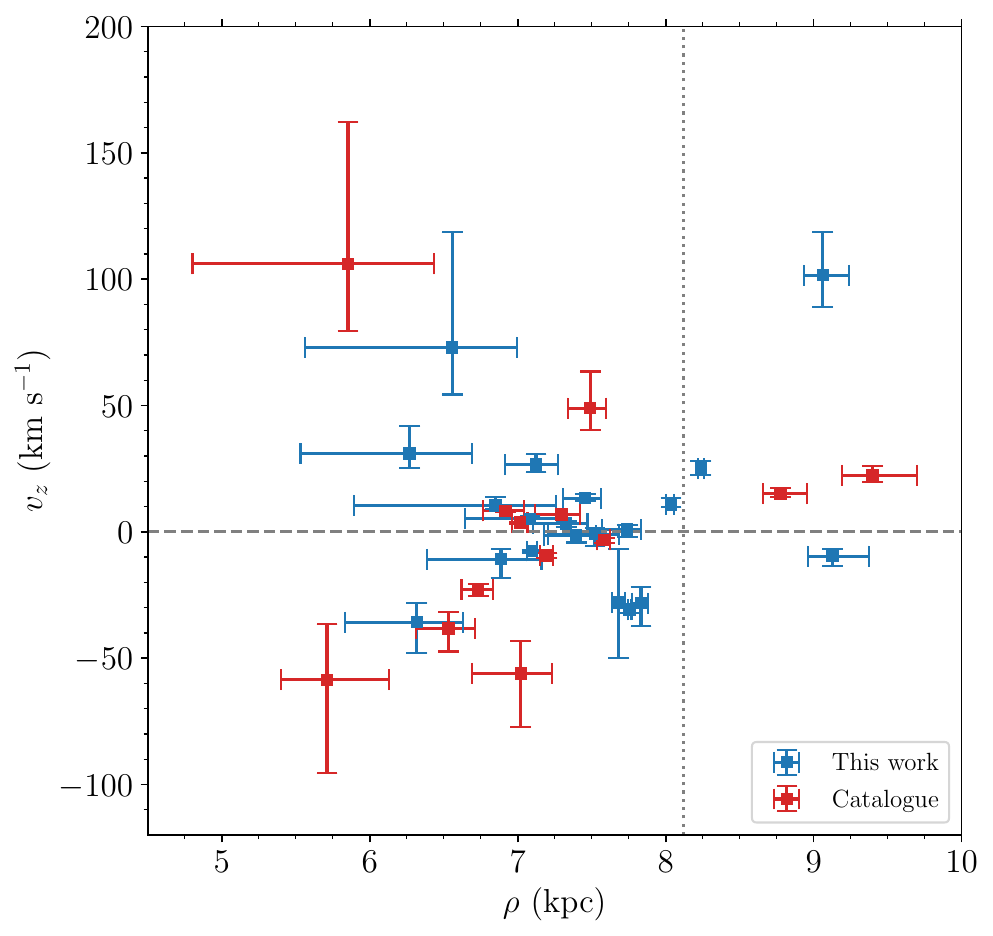}
\caption[Perpendicular component of the cylindrical Galactocentric velocity versus distance from the Galactic centre]{$z$ component of the cylindrical Galactocentric velocity ($v_{z}$) versus distance from the Galactic centre ($\rho$) for MSPs whose $z$-velocity vector are nearly perpendicular to the line of sight direction ($>70^{\circ}$). The dotted line indicates the assumed position of the Sun. Velocities calculated using MSPs in this work are shown in blue and those taken from the pulsar catalogue (\textsc{psrcat}) are shown in red. The velocity dispersion along the $z$ direction is $19(3)$ km s$^{-1}$, excluding $6$ outliers namely PSRs~J1801$-$1417, J0610$-$2100, J1911$-$1114, J1227$-$4853, J1417$-$4402 and J1431$-$4715. It is clear that near the Sun (at $\rho\approx 8.12$ kpc) $\sigma_{z}$ is small, and it becomes larger for MSPs further away from the Sun. As mentioned in the text multiple
selection effects are at play.}
\label{fig:velocity_z}
\end{figure}

We obtained cylindrical Galactocentric velocity components of $v_{\rm \rho}$ (radial), $v_{\rm \phi}$ (rotational), and $v_{\rm z}$ (perpendicular) and then calculated the components that are at least $70^{\circ}$ from the line of sight (i.e.\ for the components that are nearly perpendicular to the line of sight direction). This condition satisfies the $68\%$ confidence level condition for the measures of the velocity components. The angles with respect to the line of sight are:
\begin{equation}
    \cos(a_{\rm \rho}) = \frac{d^{2} + \rho^{2} - \sqrt{(z - z_{\odot})^{2} + R^{2}_{\odot}}}{2 \, \rho \, d},
\end{equation}
\begin{equation}
    \cos (a_{\rm z}) = \frac{z - z_{\odot}}{d},
\end{equation}
and
\begin{equation}
    \cos(a_{\rm \phi}) = \sqrt{1 - \cos(a_{\rm \rho})^{2} - \cos (a_{\rm z})^{2}},
\end{equation}
where $a_{\rm \rho}$, $a_{\rm \phi}$, and $a_{\rm z}$ are the angles between the line of sight vector and the $\rho$, $\phi$, and $z$ components, respectively. When the condition is satisfied, we have $\left | \cos a_{\rm \rho, \phi, z} \right | \leq 0.34$. We implemented the direction cosines of the line of sight vector for calculating $\cos(a_{\rm \phi})$. We have 
plotted them in Figures \ref{fig:velocity_rho}, \ref{fig:velocity_phi}, and \ref{fig:velocity_z}.

After identifying the desired components for every pulsar, we aimed to find the dispersion (the standard deviation) for each velocity component. We are not able to measure 3D space velocities of MSPs due to the lack of line-of-sight radial velocities
for many of them. We found the dispersion of the components as follows:

\begin{equation}
    \sigma_{\rho} = 101(17) \, {\rm km \, s^{-1}} ~~~~ \sigma_{\phi} = 56(9) \, {\rm km \, s^{-1}} ~~~~ \sigma_{z} = 37(5) \, {\rm km \, s^{-1}}.
\end{equation}

There are multiple outliers in the data points of the velocity components in the Figures. To remove the outliers to see any
underlying trends, we used the median absolute deviation (MAD) as a robust measure of dispersion, assuming a Gaussian velocity distribution, and followed the criteria discussed by \citet{llk+13}. Similar to  \citet{mnf+16}, for every MSP with the velocity component of $v_{\rm i}$ we obtained $\left | v_{\rm i} - M \right | / \rm MAD$, where $M$ is the global median of the i-th velocity component. To be consistent with \citet{mnf+16} analysis, we chose $\left | v_{\rm i} - M \right | / \rm MAD > 3$ to find outliers. After excluding the outliers, the dispersions of the components changed to:

\begin{equation}\label{eq:dispersions_px_cleaned}
    \sigma_{\rho} = 63(11) \, {\rm km \, s^{-1}} ~~~~ \sigma_{\phi} = 48(8) \, {\rm km \, s^{-1}} ~~~~ \sigma_{z} = 19(3) \, {\rm km \, s^{-1}}.
\end{equation}

Before interpreting the velocities we should point out that there are multiple selection effects at play in Figure  \ref{fig:velocity_z}. This figure tends to imply that $\sigma_{z}$ of nearby MSPs is much lower compared to that of further MSPs. 
This means it is more likely to find MSPs with low $v_{z}$ near the Sun. MSPs that are near us are either moving very slowly and will always be near us, or they are moving fast and just happen to be passing near us. 
MSPs rising out of the Galactic plane will slow down eventually (e.g.\ a pulsar with a vertical velocity of $\approx 80$ km\,s$^{-1}$ will rise $\approx 2$ kpc) and then return back towards the plane. The amount of time such pulsars spend around $2$ kpc is much more than the time they spend nearby the Sun when they are passing quickly through the plane. Therefore, for a short amount of time they are expected to be observed with high luminosity, and so it is easier to measure their parallaxes and proper motions: this leads to a strong selection effect in the sample. Another selection effect is that the more distant pulsars in the sample are likely to be drawn from the luminous end of the pulsar luminosity function and could be seen at greater $z$-distances.
Finally, the proper motions of fast moving pulsars are easier to determine than those of slow moving pulsars, which will likely boost the mean tangential velocity of the sample
at larger distances. All these selection effects are at play, and simple comparisons of the observed MSP velocities as a function of $z$ or distance needs to be performed with caution
and best done by simulations as was attempted for the slow pulsars by \cite{lbh97}.

\subsection{Asymmetric drift in the Galaxy}
An asymmetric drift is predicted to be observed in normal pulsars with characteristic ages greater than $10^{7}$ yr \citep{hp97}. MSPs have characteristic ages greater than $10^{8}$ yr and are expected to be old enough to have reached a virialized state; so they are also expected to demonstrate such a drift. \citet{cc97} predicted an asymmetric drift of $13 \, {\rm km \, s^{-1}}$ for the case of a uniform surface density model, and $25 \, {\rm km \, s^{-1}}$ for the case of an exponential surface density model, both with the MSP velocity dispersion of $60 \, {\rm km \, s^{-1}}$. In Figure \ref{fig:velocity_phi}, we can clearly see most of the MSPs moving opposite to the direction of Galactic rotation, which is the signature of the asymmetric drift. Utilising the rotational velocities, $V_\phi$, 
we obtained a median rotational velocity of $38(16) \, {\rm km \, s^{-1}}$ and the mean rotational velocity of $45(13) \, {\rm km \, s^{-1}}$. By removing the outliers (as defined in \ref{sec:velocity_dispersion}), we obtained the median value of $33(14) \, {\rm km \, s^{-1}}$ and the mean value of $38(11) \, {\rm km \, s^{-1}}$. Our values are higher than the $25(12) \, {\rm km \, s^{-1}}$ obtained by \citet{tsb+99}. Our median value is closer to $25 \, {\rm km \, s^{-1}}$, the predicted value corresponding to the exponential surface density model of \citet{cc97} although the MSPs in our sample have higher mean velocities than theirs.

\section{Comparison with previous work}\label{sec:discussions}
\subsection{Millisecond Pulsars vs. Normal Pulsars}

Soon after their discovery, a population study by \citet{go70}, based on data from just $41$ young pulsars, argued that pulsars are probably born in a disk distribution with an initial scale height of $\sim 80$ pc (similar to the normal OB stars), and move with average velocities of $\sim 100 \, {\rm km \, s^{-1}}$. In the first large scale study of pulsar proper motions, \citet{las82} studied a sample of $26$ and showed that the pulsars with Galactic scale height greater than a few tens of parsecs tend to be moving away from the Galactic plane, probably due to a velocity kick at birth. They found pulsars had an rms velocity of about $210$ km s$^{-1}$. Later, \cite{ll94}
studied the youngest pulsars with proper motions, and found they had a mean
space velocity at birth of $450(90)$ km s$^{-1}$.

By modelling the kinematics of the spatial distribution of MSPs, \citet{cc97} estimated that such pulsars on average receive a $z$-velocity kick at birth of just $52^{+17}_{-11} \, {\rm km \, s^{-1}}$, much less that than of the normal pulsars. They suggested that the rms speed of young pulsars is $\sim 5$ time larger than that of MSPs, and a significant contribution to the observed $z$-velocity (the velocity component along Galactocentric $z$) of MSPs  originates from the diffusive processes that affect all
old stars in the disk. 

\begin{figure}
\centering
 \includegraphics[width=0.5\textwidth]{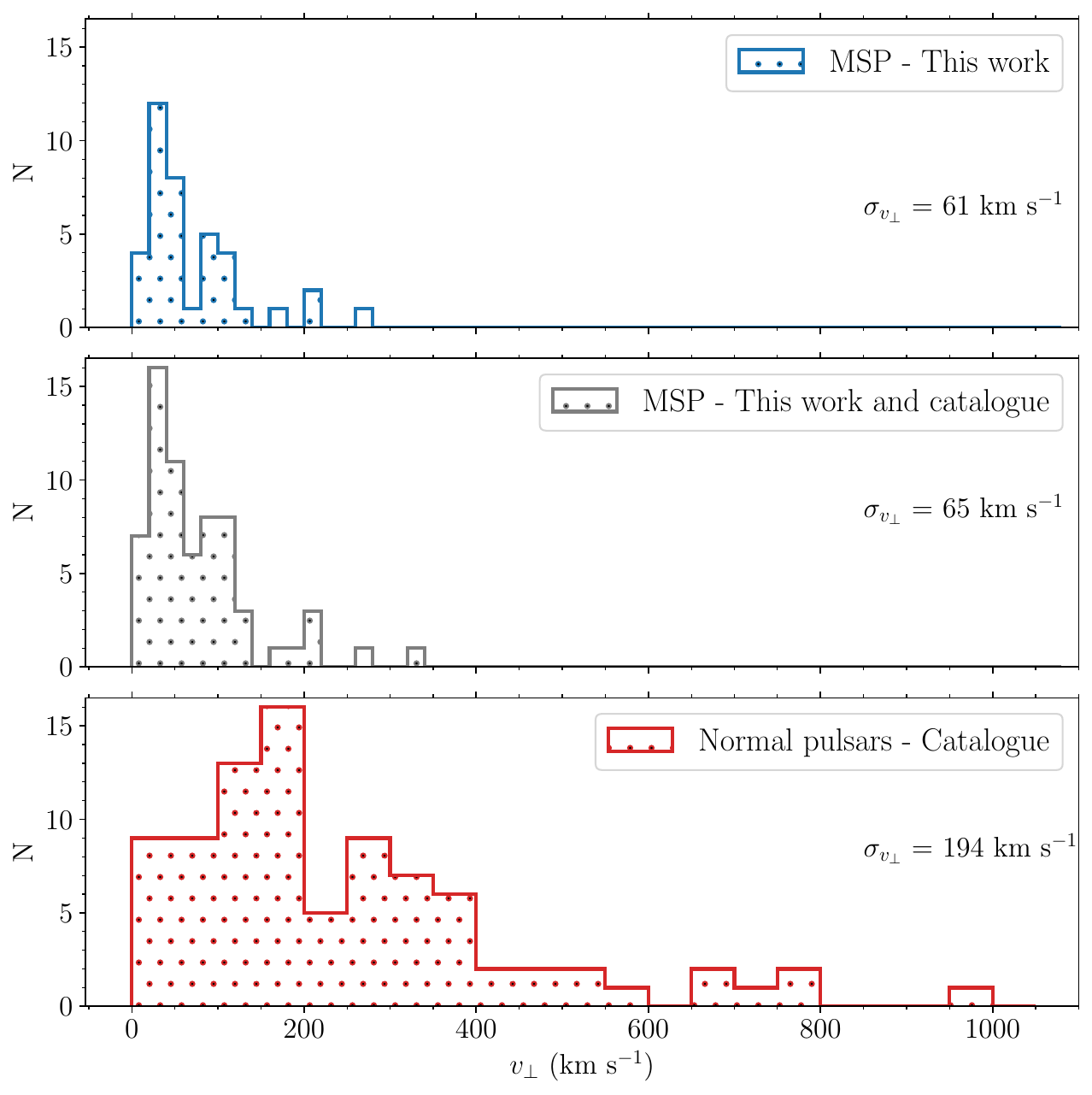}
\caption[Histograms of transverse velocities of MSPs and normal pulsars]{Histograms of transverse velocities for our MSP sample (top panel and in blue), our MSPs and MSPs in the pulsar catalogue (middle panel and in gray), and the normal pulsars in pulsar catalogue that have $>3\sigma$ significance in parallaxes (bottom panel and in red). This plot indicates how different the distributions of MSP velocities and the normal pulsar velocities are.}
\label{fig:vel_hist_msp_normal}
\end{figure}

We plotted the histograms of the transverse velocities of the normal pulsars from the pulsar catalogue and the MSPs in Figure \ref{fig:vel_hist_msp_normal}, to highlight how different the two distributions are. In this Figure, the top, middle, and bottom panels show the distributions of MSPs from this work, MSPs from the combination of this work and the pulsar catalogue, and normal pulsars from the pulsar catalogue, respectively. For the $87$ normal pulsars in the pulsar catalogue with significant parallaxes and proper motions ($>3\sigma$), we found a mean velocity of $246(21) \, {\rm km \, s^{-1}}$. Comparing the mean velocities showed that the normal pulsars seem to be faster than MSPs by a factor of $\sim3.2$. The characteristic age of MSPs is generally higher than that of normal pulsars, and accordingly, the old pulsars seem to be slower than young pulsars. 
If we restrict our attention to the pulsars with characteristic
ages less than $3$ Myr, we find a mean transverse velocity 
of $283(44)$ km s$^{-1}$, $\sim3.7$ times faster than the MSPs.

\begin{figure}
\centering
 \includegraphics[width=0.5\textwidth]{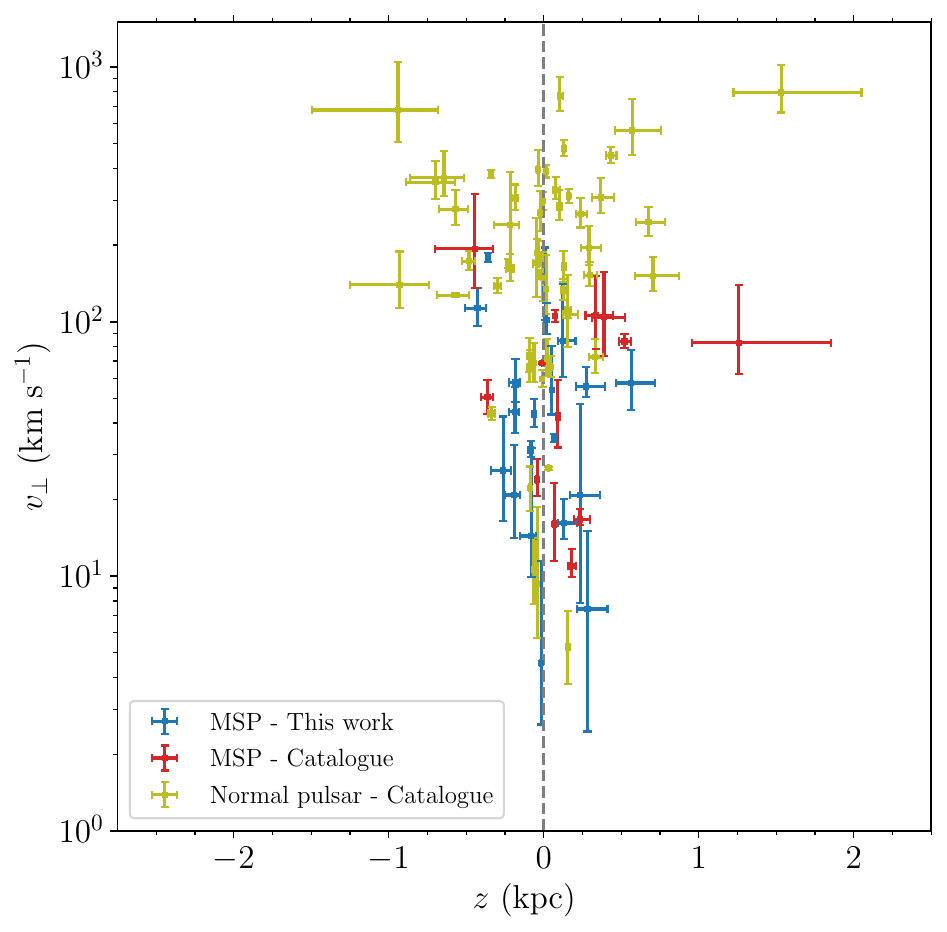}
\caption[Transverse velocities versus Galactocentric $z$ coordinate of pulsars]{Transverse velocities versus Galactocentric $z$ coordinate of pulsars. This plots shows the difference between the velocities of MSPs and normal pulsars. Velocities of MSPs from this work, MSPs from the pulsar catalogue, and normal pulsars from pulsar catalogue \citep{mht+05} are shown in blue, red, and olive, respectively. Pulsars from the pulsar catalogue are shown if
they have at least $3\sigma$ significance in parallax. The mean velocity of the normal pulsars is $246(21)$ km s$^{-1}$, $\sim3.2$ times the mean MSP transverse velocity of $77(8)$ km s$^{-1}$. Also, the velocity dispersion of the normal pulsars is $194(15)$ km s$^{-1}$, $\sim3$ times the MSP velocity dispersion of $65(6)$ km s$^{-1}$.
Pulsars with velocities $< 100$ km s$^{-1}$ are often closer to the Galactic plane. }
\label{fig:vel_glat}
\end{figure}

The velocities of the MSPs and normal pulsars as a function of Galactocentric $z$ are shown in Figure \ref{fig:vel_glat}. This figure demonstrates that pulsars with velocities $<100$ km s$^{-1}$ are more concentrated in the Galactic plane and are less scattered compared to the pulsars with velocities $>100$ km s$^{-1}$. In addition, on average, normal pulsars seem to have a wider $z$-distribution around the Galactic plane (i.e.\ larger $z$ scale height) compared to the MSPs
consistent with their higher velocities.
This figure provides additional evidence for the results from the numerical studies of \citet[][and references therein]{bv91}; \citet{tb96,cc97} that suggested MSPs have lower velocities compared to the young, long-period pulsars. The higher velocities of normal pulsars reinforce the idea that the high recoil velocity that pulsars receive during their birth in supernova explosions \citep{bur13,vic17,dgb+19}, and are in agreement with some of the core-collapse supernovae simulations performed by \citet{wjm12,mul20}. On the other hand, some of the low velocities of normal pulsars might be due to the weak supernova kicks for a sub-population of the pulsars \citep{wmt+21}. 
Some caution in interpreting Figure \ref{fig:vel_glat} is required, as
many selection effects are at play.

\subsection{Velocity dispersions for millisecond pulsars}
The dispersions of the $3$ velocity components that \citet{mnf+16} obtained from the analysis of NANOGrav nine-year MSP timing data set, after excluding outliers, are $\sigma_{\rm \rho} = 46 \, {\rm km \, s^{-1}}$, $\sigma_{\rm \phi} = 40 \, {\rm km \, s^{-1}}$, and $\sigma_{\rm z} = 24 \, {\rm km \, s^{-1}}$. We obtained $\sigma_{\rm \rho} = 63(11) \, {\rm km \, s^{-1}}$, $\sigma_{\rm \phi} = 48(8) \, {\rm km \, s^{-1}}$, and $\sigma_{\rm z} = 19(3) \, {\rm km \, s^{-1}}$. Our $\sigma_{\rm \phi}$ and $\sigma_{\rm z}$ are thus comparable with their values. Also, they calculated the dispersion of velocity components, using the fit equations to the local stellar data provided by \citet{ab09}, to be $\sigma_{\rm \rho} = 34 \, {\rm km \, s^{-1}}$, $\sigma_{\rm \phi} = 22 \, {\rm km \, s^{-1}}$, and $\sigma_{\rm z} = 18 \, {\rm km \, s^{-1}}$ for the characteristic age of $\tau \sim 5$ Gyr, and $\sigma_{\rm \rho} = 42 \, {\rm km \, s^{-1}}$, $\sigma_{\rm \phi} = 28 \, {\rm km \, s^{-1}}$, and $\sigma_{\rm z} = 24 \, {\rm km \, s^{-1}}$ for $\tau \sim 10$ Gyr. This is showing that our $\sigma_{\rm z}$ is consistent with the model fit corresponding to the characteristic age of $\tau \sim 5$ Gyr, but our $\sigma_{\rm \rho}$ and $\sigma_{\rm \phi}$ are closer to the model fit corresponding to the characteristic age of $\tau \sim 10$ Gyr. Both comparisons show that the MSPs are consistent with having been drawn from the old disk stellar population, and that they are subject to kick velocities at birth, as discussed in the next section.

The fact that the $z$-components are the lowest is perhaps easiest
to understand. MSPs with low $z$-velocities spend more time
near that galactic disk (and hence the Sun) and are
preferentially detected by MSP surveys, plus as mentioned
before, the $z$-velocity is on average lower than
the birth $z$-velocity as it exhibits simple harmonic
motion in the Galactic potential.

\begin{figure}
\centering
 \includegraphics[width=0.5\textwidth]{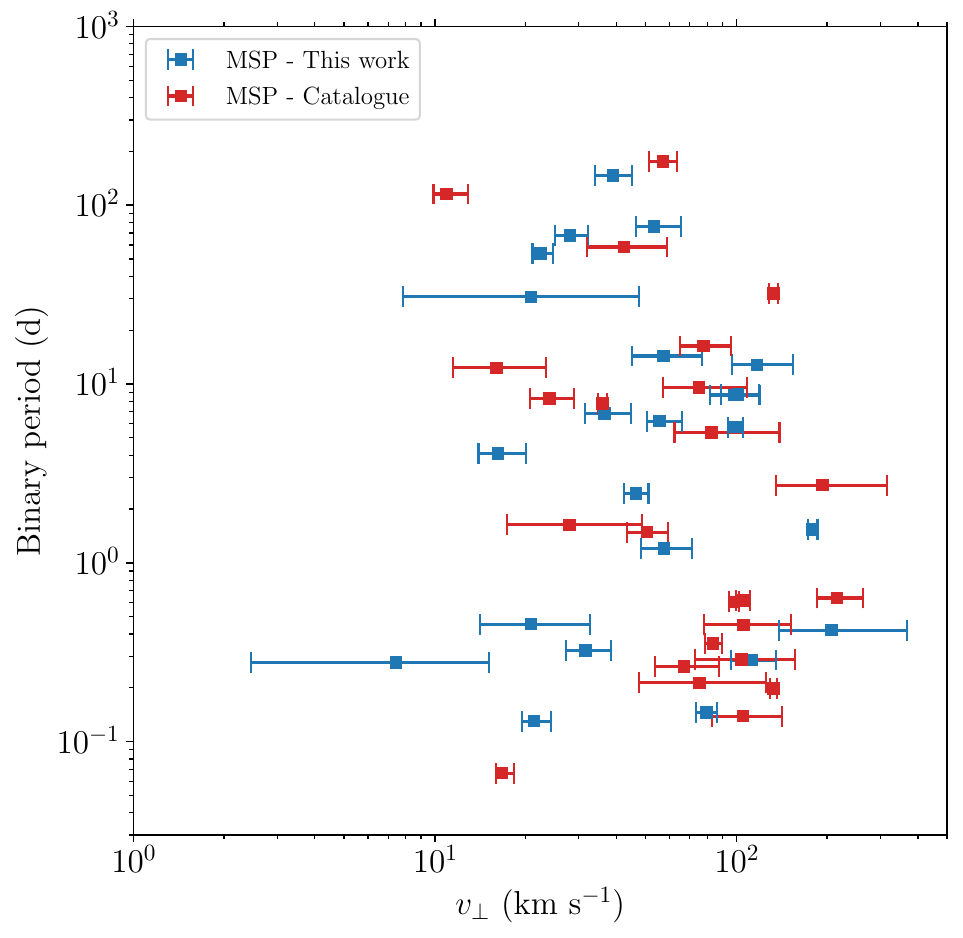}
\caption[Transverse velocity versus orbital period for binary MSPs]{Transverse velocity versus orbital period for binary MSPs. Velocities calculated using MSPs in this work are in blue and from the pulsar catalogue are in red. The correlation coefficient between velocity and binary period (excluding PSRs~J1024$-$0719, J1300$+$1240, and J2234$+$0611 for reasons discussed in the text) was found to be $-0.27$, indicating a weak anti-correlation,
similar to that predicted.}
\label{fig:velocity_vt_pb}
\end{figure}

\subsection{Velocity distributions for millisecond pulsars}
\citet{hll+05} studied the kinematics of 233 pulsars and found the mean two-dimensional speeds of $211(18) \, {\rm km \, s^{-1}}$ for all pulsars in their sample, $307(47) \, {\rm km \, s^{-1}}$ for pulsars with characteristic ages less than $3$ Myr, $87(13) \, {\rm km \, s^{-1}}$ for recycled pulsars, and $209(19) \, {\rm km \, s^{-1}}$ for the normal pulsars with characteristic ages greater than $3$ Myr. Our mean velocity of $78(8) \, {\rm km \, s^{-1}}$ for MSPs is consistent with their mean two-dimensional speed of $87(13) \, {\rm km \, s^{-1}}$ for recycled pulsars. Also, \citet{hll+05} obtained the mean 2D speeds of $77(16) \, {\rm km \, s^{-1}}$ for seven isolated MSPs and $89(15) \, {\rm km \, s^{-1}}$ for 28 binary MSPs. These are comparable with the mean velocities of $67(12) \, {\rm km \, s^{-1}}$ and $77(9) \, {\rm km \, s^{-1}}$ for the $16$ solitary and $49$ binary MSPs in our sample, respectively, excluding PSR~J1024$-$0719. \citet{jnk98}, by studying scintillation parameters for a sample of 49 pulsars, suggested that binary MSPs have higher velocities compared with the isolated MSPs. \citet{tsb+99} calculated velocities for a sample of $23$ MSPs and presented that, on average, the binary MSPs are one-third faster than isolated MSPs. However, \citet{hll+05} reported that there is no significant difference between the mean velocities of both. Our results are also showing that the mean velocities of both are not significantly different. 

\begin{figure}
\centering
 \includegraphics[width=0.5\textwidth]{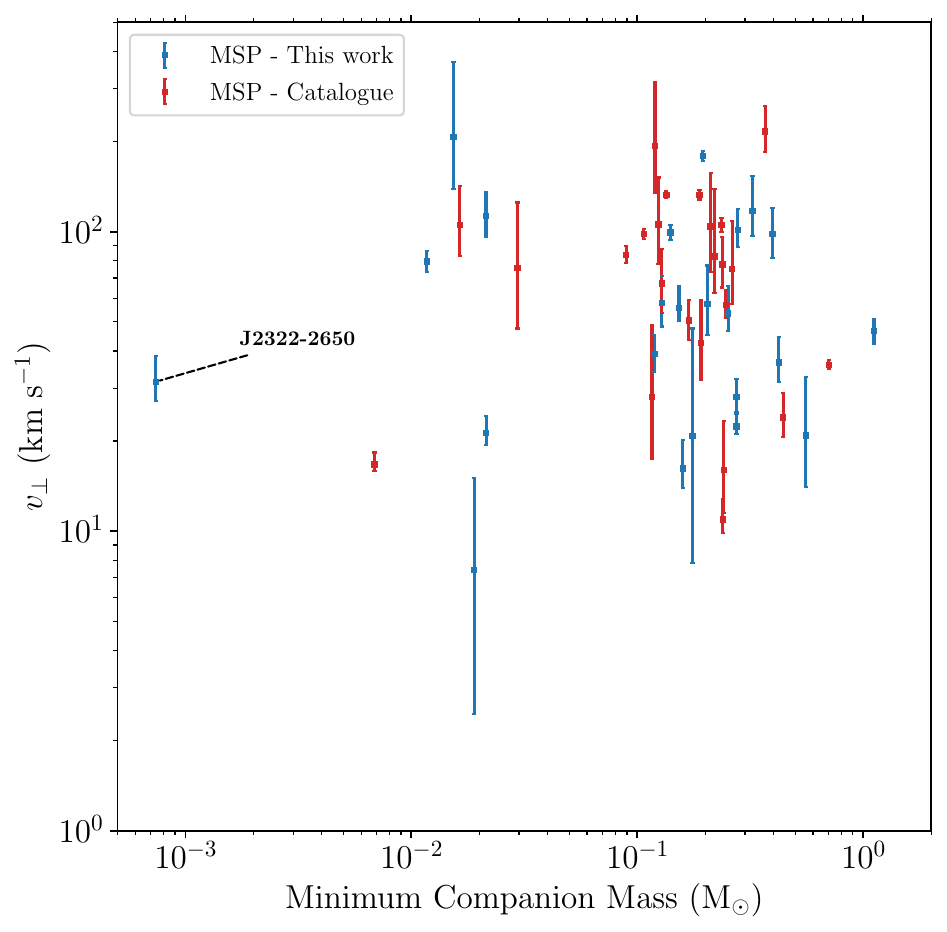}
\caption[Transverse velocity versus minimum companion mass for binary MSPs]{Transverse velocity versus minimum companion mass assuming orbital inclination of $90^{\circ}$ and the pulsar mass of $1.35$ M$_{\odot}$ for binary MSPs. Velocities calculated using MSPs in this work are in blue and from the pulsar catalogue are in red. The correlation coefficient between velocity and minimum companion mass (excluding PSRs~J1024$-$0719, J1300$+$1240, and J2234$+$0611 for reasons discussed in the text) was found to be $-0.10$, indicating almost no correlation.}
\label{fig:velocity_vt_minmass}
\end{figure}

Using simulations, \citet{tb96} predicted a mild inverse correlation between the recoil velocity (the post-explosion velocity that a pulsar receives at birth) and the orbital period of binary MSPs, depending upon the birth component masses of the binary, the separation between the two objects, and the evolution of the system during the common-envelope and mass transfer stages. 
\citet{tsb+99} did not find any significant correlation with orbital period using a sample of $23$ binary MSPs. In addition, \citet{hll+05} looked for this correlation using the binary MSPs in their pulsar sample, but they did not find any either. We also investigated the relationship between transverse velocity and orbital period of binary MSPs in Figure \ref{fig:velocity_vt_pb}, and measured the correlation coefficient to be $-0.24$ (excluding the very long orbital period pulsar PSRs~J1024$-$0719, the planet pulsar J1300$+$1240, and the eccentric MSP J2234$+$0611 - which probably has a complex evolutionary history) showing a weak anti-correlation. 
We need to keep in mind that many selection effects are influencing our observed MSP
distribution. For instance, 
binary MSPs with short orbital periods and heavy companions 
might be less likely to appear in pulsar surveys unless acceleration
searches have been undertaken. 
Nevertheless, \cite{tb96} predicted an approximate velocity range of $25$--$270$ km s$^{-1}$ for binary MSPs and that those above $300$ km s$^{-1}$ should
be very rare. This is borne out by our observations of our sample's transverse velocities.
In the standard model MSPs obtain their velocities as the vector sum of three
distinct mechanisms.
First, MSP progenitors, the binaries containing OB stars, are formed in large stellar nurseries
where they obtain velocities of 10s of km s$^{-1}$ due to
stellar interactions. Second, when the neutron star is produced
there is a Blaauw momentum kick to the binary as a 
result of the mass loss of the supernova \citep{bla61},
and possibly an asymmetric kick imparted to the neutron star.
To get very high velocities requires very compact binaries,
large mass loss and asymmetric retrograde kicks that do not
disrupt the orbit. These must be rare in MSP formation.
As we only measure the 2D velocities the presence of some
very low velocities is not unexpected and could be a projection effect, but MSP velocities
can help constrain models for neutron star formation in population synthesis
codes 
such as COMPAS \citep{rab+22} and STARTRACK \citep{bkb02}.

Any correlation between the minimum masses of companions of binary MSPs and their transverse velocities of binary MSPs was explored. We used the pulsar catalogue to find the minimum mass of the companions and plotted them as a function of transverse velocity in Figure \ref{fig:velocity_vt_minmass}. The minimum masses were calculated assuming the orbital inclination angle to be $90^{\circ}$ and the mass of the pulsar to be $1.35$ M$_{\odot}$. We measured the correlation coefficient between the minimum mass and the transverse velocity (excluding PSRs~J1024$-$0719, J1300$+$1240, and J2234$+$0611) to be only $-0.10$, showing at best
a very weak anti-correlation.

\section{Conclusions}\label{sec:conclusions}

We have undertaken a study on the parallaxes and proper motions of $77$ MSPs, observed with the MeerKAT radio telescope, over a time span of $\sim 3$ years. Out of the $77$ MSPs, $35$ and $69$ had significant parallaxes and proper motions (above $3\sigma$), respectively. We calculated the transverse velocities for MSPs that had both significant parallaxes and proper motions. For the MSPs for which we did not have significant parallaxes, we used parallaxes taken from the literature. Pulsars with significant parallaxes and proper motions from the pulsar catalogue \textsc{psrcat} were also added to our MSP sample to produce the largest MSP velocity sample to date. We found that the transverse velocity of MSPs has a mean of $78(8)$ km s$^{-1}$ and an rms scatter of $101$ km s$^{-1}$. 
    
The Lutz-Kelker effect on the derived transverse velocities was investigated for a sample of $35$ pulsars with parallaxes. The median transverse velocities increased by $\sim 2\%$ and $\sim 13\%$ for parallaxes with $>10\sigma$ and $<10\sigma$ significance, respectively. The median of the whole population increased by $\sim 7\%$.
    
Out of our $69$ MSPs with significant ($>3\sigma$) proper motions, the long orbital period pulsar J1024$-$0719 had the highest transverse velocity of $278^{+44}_{-32}$ km s$^{-1}$, and the $12.4$ d orbital period binary pulsar PSR~J1652$-$4838 had the lowest transverse velocity of just $5^{+7}_{-2}$ km s$^{-1}$.

Although the mean velocity of $72(8)$ km s$^{-1}$ for $49$ binary MSPs is slightly faster than that of $67(12)$ km s$^{-1}$ for $16$ isolated ones, there was no evidence that their distributions are statistically significantly different.
In comparison to the normal pulsars in the pulsar catalogue with the mean transverse velocity of $246(21)$ km s$^{-1}$, MSPs had a much lower mean transverse velocity of $78(8)$ km s$^{-1}$.
    
The velocity dispersions of the pulsars in the (cylindrical) Galactocentric velocity components (radial, rotational, and perpendicular) were measured to be $\sigma_{\rm \rho} = 63(11) \, {\rm km \, s^{-1}}$, $\sigma_{\rm \phi} = 48(8) \, {\rm km \, s^{-1}}$, and $\sigma_{\rm z} = 19(3) \, {\rm km \, s^{-1}}$ after removal of a small number of outliers. The lower $z$-velocity component is likely a consequence of the fact that high velocity MSPs will spend less time near the Sun suppressing their representation in pulsar surveys. 
    
The expected asymmetric drift was clearly seen in the rotational component of the velocities, and the mean value was found to be $38(11) \, \rm km~s^{-1}$ after removing outliers. This is consistent with the number predicted by \citet{cc97}, who model the formation of MSPs in our galaxy. They predicted an asymmetric drift of approximately $25$ km s$^{-1}$ for their more realistic (exponential surface density) Milky Way model.

The substantially increased sample of MSPs with good parallaxes and proper motions presented in this study bolsters the case that MSPs arise out of the old disk of the Milky Way and, are subject to kick velocities at birth significantly smaller than those seen for young pulsars \citep{hll+05}. We explored the distribution of MSP velocities versus orbital period and found a weak anti-correlation, in agreement with the predictions of \citet{tb96}, based on simulations of the recoil velocities of MSPs in various binary stellar evolutionary scenarios.

\section*{Acknowledgements}

We thank Dr. Simon Stevenson and Dr. Hao Ding for their suggestions and comments that improved the manuscript. The MeerKAT telescope is operated by the South African Radio Astronomy Observatory, which is a facility of the National Research Foundation, an agency of the Department of Science and Innovation. 
MS, MB, CF, MTM, DJR, and RMS acknowledge support through the Australian Research Council Centre of Excellence for Gravitational Wave Discovery (OzGrav), through project number CE17010004.
RMS acknowledges support through Australian Research Council Future Fellowship FT190100155.
MK acknowledges significant support from the Max-Planck Society (MPG) and the MPIfR contribution to the PTUSE hardware.
This work used the OzSTAR national facility at Swinburne University of Technology. OzSTAR is funded by Swinburne University of Technology and the National Collaborative Research Infrastructure Strategy (NCRIS) that also supports the pulsars.org.au data portal
that was used extensively for this work.
This research has made use of NASA's Astrophysics Data System and software such as: \textsc{psrchive} \citep{vdo12}, \textsc{tempo2} \citep{hem06,ehm06}, \textsc{temponest} \citep{lah+14}, \textsc{psrcat} \citep{mht+05}, \textsc{pulseportraiture} \citep{pdr14,p19}, \textsc{pygedm} \citep{pfd21}, \textsc{numpy} \citep{numpy}, \textsc{scipy} \citep{scipy}, \textsc{matplotlib} \citep{matplotlib}, \textsc{ipython} \citep{ipython}, \textsc{astropy} \citep{astropy,astropy_v2}.

\section*{Data Availability}
Data is available from the Swinburne pulsar portal: https://pulsars.org.au
 


\bibliographystyle{mnras}
\bibliography{main} 




\appendix

\section{On the use of Galactic electron density models for MSP distances and velocities}\label{sec:dist_comparison_dist_ratio}

The integrated electron column density ($n_{\rm e}$) along the line of sight from the Earth to a pulsar at distance $d$ relates to the DM of the pulsar as
\begin{equation}
        {\mathrm {DM}} = \int_{0}^{d} n_{\mathrm{e}}(l) dl.
\end{equation}
The distance to a pulsar can be derived from an observational DM by having a model which describes the distribution of electrons throughout the Galaxy. An inaccurate electron-density model might result in overestimating or underestimating distances.  
By measuring new pulsar distances via pulsar parallaxes, as in this work, we are able to provide an improved basis for refining the ISM models for different lines of sight. There are two widely used DM models, namely NE2001 \citep{cl02,cl03} and YMW16 \citep{ymw17}. The comparison of the two has been carried out by many authors \cite[e.g., ][]{ymw17,dgb+19,occ20,pfd21}. They found
that the different models have performance that is dependent on where one is looking on the sky, and thus one model works better than others depending on which part of the Galactic ISM is being probed by any given pulsar.

\subsection{Distances}

Using the two Galactic electron-density models: NE2001 and YMW16, implemented in \textsc{PyGEDM} \citep{pfd21}, we have calculated DM-based distances of the $35$ MSPs with $>3\sigma$ significance in their parallaxes. In addition, for MSPs with $<3\sigma$ significance in their parallaxes, we added those MSPs whose DM distances were less than the lower distance limits obtained from the probability distributions (the third column in Table \ref{tab:parallax_distance}). We also added other MSPs from the pulsar catalogue \citep{mht+05} with $>3\sigma$ significance in parallax into our MSP sample and calculated their DM-based distances. We excluded all MSPs in globular clusters as they have different origins and their velocities are
dominated by their host clusters.

\begin{figure}
\centering
 \includegraphics[width=0.5\textwidth]{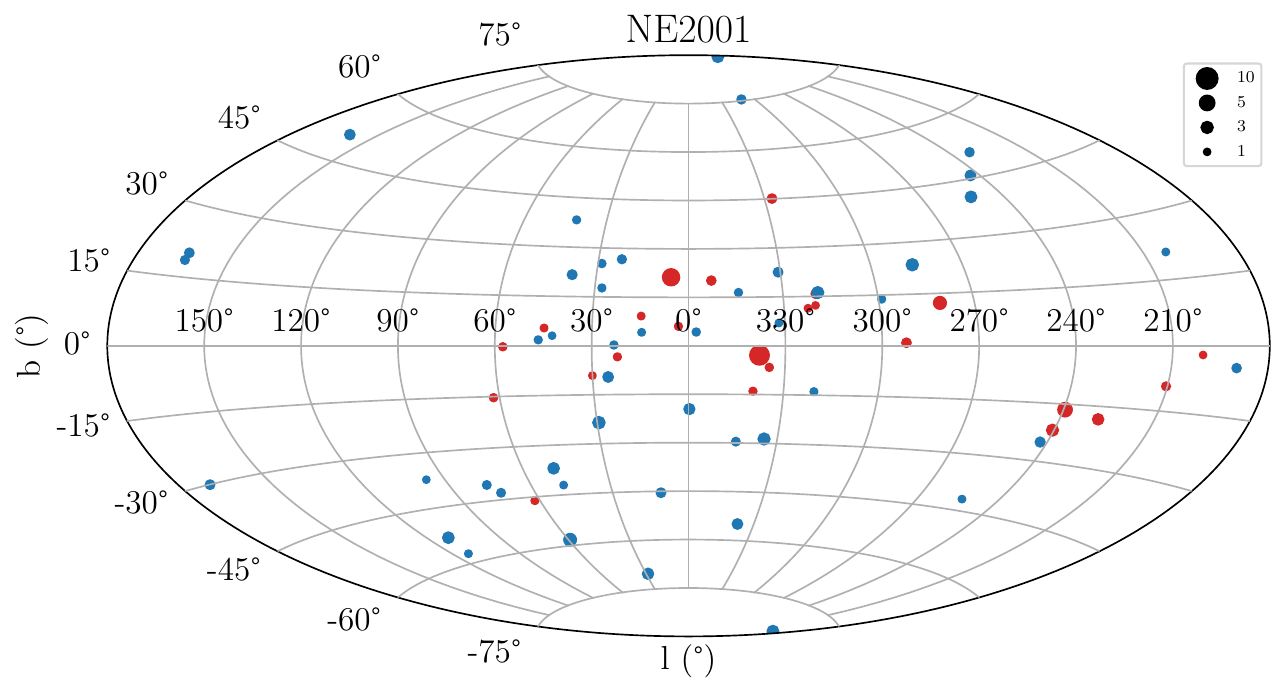}
 \includegraphics[width=0.5\textwidth]{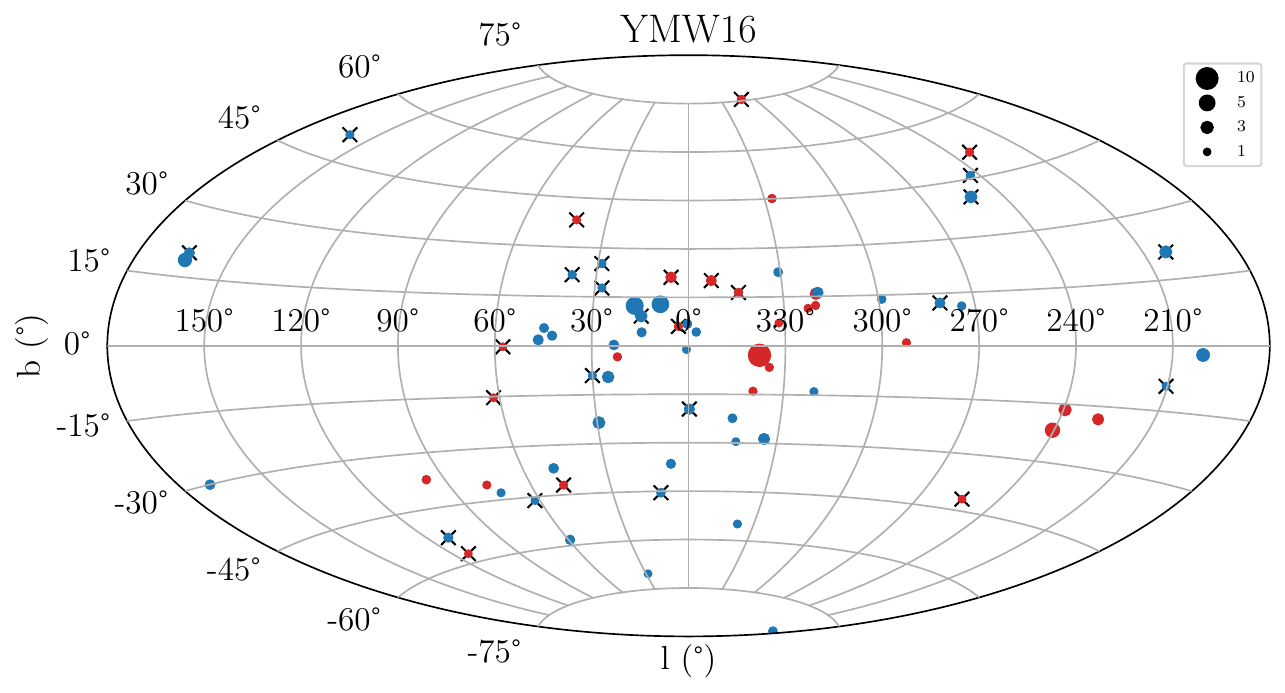}
\caption[Distance ratios for the NE2001 and YMW16 distance models and parallaxes]{The ratios of the parallax derived distances and DM derived distances. Top panel: distance ratios for the NE2001 model. Bottom panel: Distance ratios for the YMW16 model. The blue points indicate DM model distance underestimation ($d_{\varpi}/d_{\rm DM}$), and the red points indicate distance overestimation ($d_{\rm DM}/d_{\varpi}$). The sizes of the black circles in the legend indicate the distance ratios. The black crosses in bottom panel show the parallaxes used for establishing the YMW16 model
that we might expect to fit well.
}
\label{fig:dm_px_distance_ratio}
\end{figure}

The top and bottom panels of Figure \ref{fig:dm_px_distance_ratio} show the comparison of parallax distances with DM distances obtained from the NE2001 and YMW16 models, respectively. In this Figure, points in red show a DM distance overestimation (i.e.\ MSPs are closer than what the models predict) and points in blue show a DM distance underestimation (i.e.\ MSPs are further than what the models predict). The ratios of $d_{\rm DM} / d_{\rm \varpi}$ and $d_{\rm \varpi} / d_{\rm DM}$ are represented by the area of the circles in red and blue for the cases of overestimation and underestimation, respectively. The black circles in the legend of Figure \ref{fig:dm_px_distance_ratio} provide the standard circle sizes for the ratios of $10$, $5$, $2$, and $1$. We found the maximum underestimation based on the YMW16 model for PSR~J1737$-$0811 with the distance ratio of $d_{\varpi} / d_{\rm DM} \approx 5.9$ (considering the lower limit of its parallax distance), and the maximum overestimation was for PSR~J1652$-$4838 with the distance ratio of $d_{\rm DM} / d_{\varpi} \approx 10.5$.

The $16\%$, $50\%$, and $84\%$ of the underestimation and overestimation of the YMW16 distances ratios were obtained to be $1.4^{+1.1}_{-0.3}$ and $1.2^{+1.1}_{-0.1}$, respectively. In addition, the underestimation and overestimation of the NE2001 distances were obtained to be $1.6^{+1.1}_{-0.5}$ and $1.3^{+1.8}_{-0.2}$, respectively. Also, the distance ratios of $d_{\varpi} / d_{\rm DM}$ were obtained for YMW16 model to be $1.1^{+0.7}_{-0.3}$ and for NE2001 to be $1.2^{+1.3}_{-0.4}$. The comparison of distance ratios highlighted that there are obviously patches in Figure \ref{fig:dm_px_distance_ratio} that the DM-based distances are systematically overestimated or underestimated by the two models.

\subsection{Velocities}
\label{sec:transverse-velocities}

\begin{figure}
\centering
 \includegraphics[width=0.5\textwidth]{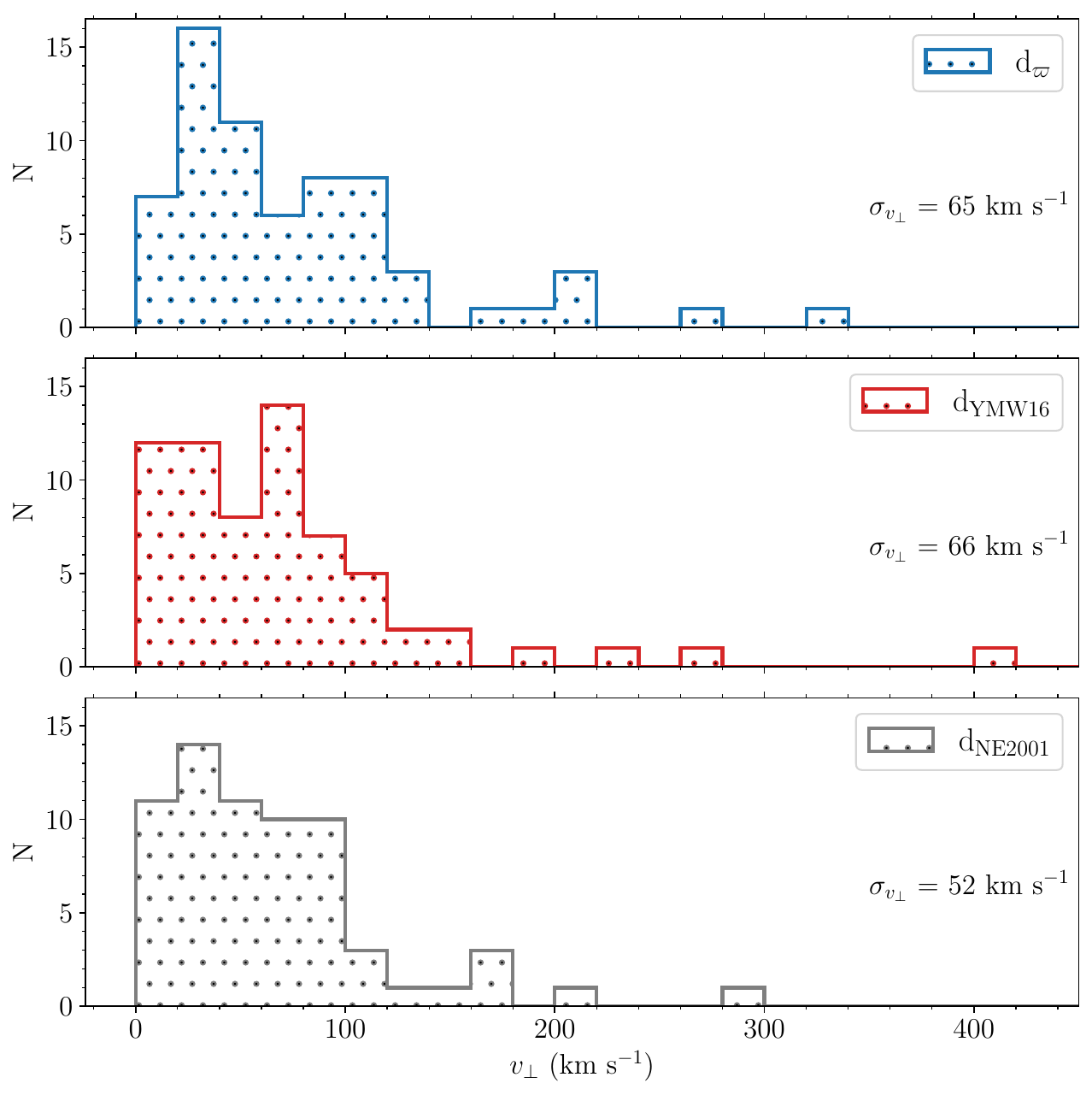}
\caption[Histograms of parallax-based velocities compared to the DM-based velocities]{Histograms of parallax-based velocities compared to the DM-based velocities. The transverse velocities are calculated using distances derived from the parallax measurements in the top panel, YMW16 DM model in the middle panel, and NE2001 DM model in the bottom panel. As discussed in Section \ref{sec:transverse-velocities}, there is an excess of pulsars with very low velocities (lower two panels) for which distances have been derived from an ISM model, compared to the sample compiled from pulsars with parallaxes only (upper panel). This is an artifact of these models placing pulsars too close to the Sun. The velocity standard deviation of $\sigma_{\perp}$ for the YMW16 model compared to the one for NE2001 model shows the number of low velocities in NE2001 model are higher.
\label{fig:vel_hist_px_dm_dist}}
\end{figure}

The top panel of Figure \ref{fig:vel_hist_px_dm_dist} presents the histogram of transverse velocities for the total sample that possesses parallax measurements. The mean transverse velocity is $78(8) \, {\rm km \, s^{-1}}$. The fastest moving MSP for these parallax-based velocities is the planet pulsar PSR~J1300$+$1240 \citep{wol90a}, which has a velocity of $330^{+20}_{-18} \, {\rm km \, s^{-1}}$, and the slowest-moving, PSR~J1652$-$4838, has a velocity of just $5^{+7}_{-2} \, {\rm km \, s^{-1}}$. 
Using the code \footnote{\url{https://github.com/jantoniadis/GAIA_pulsars_eDR3/blob/main/GAIA_pulsars_eDR3.ipynb}} provided by \citet{ant21}, we explored whether the LK corrections might affect our velocities by breaking our
sample into two groups, those with $>10\sigma$ parallaxes and those with lower significance.
The median velocities of the first group before and after the LK correction were $67$ and $68$ km s$^{-1}$ respectively
and the second group $43$ and $49$ km s$^{-1}$. Clearly, the LK correction raises the velocities of the second group
and brings their median closer to that of the (more accurate) group, as might be predicted by the LK selection effect. Overall,
the median velocities of the entire population before and after the LK correction were
$46$ and $50$ km s$^{-1}$, an increase of $\sim 7$ \%.


To illustrate the effect of using DM-based distances rather than our preferred parallax-based distances, we have derived transverse velocities using DM-based distances obtained from NE2001 and YMW16 models. The middle and the bottom panels of Figure \ref{fig:vel_hist_px_dm_dist} present the histograms of transverse velocities for the MSPs that their velocities are calculated from YMW16 and NE2001 models, respectively. 

Using the YMW16 model, the mean transverse velocity is $71(8) \, {\rm km \, s^{-1}}$. The fastest moving MSP, PSR~J1300$+$1240, would have a velocity of $407(100) \, {\rm km \, s^{-1}}$ in the DM model, and the slowest-moving, PSR~J1446$-$4701, has a velocity of $6^{+6}_{-4} \, {\rm km \, s^{-1}}$, assuming $25\%$ distance errors. Using the NE2001 model, the mean transverse velocity is $65(6) \, {\rm km \, s^{-1}}$. The fastest moving MSP in this model, PSR~J0610$-$2100, has a velocity of $292(80) \, {\rm km \, s^{-1}}$, and the slowest-moving, PSR~J1446$-$4701, has a velocity of $6^{+6}_{-4} \, {\rm km \, s^{-1}}$, again assuming $25\%$ distance errors.

It is curious to note that distance errors in YMW16 are responsible for a marked increase in the number of very low velocity MSPs
(see Figure \ref{fig:vel_hist_px_dm_dist}, lower two panels).
In reality there are much fewer low velocity MSPs in the sample based on parallax distances only (Figure \ref{fig:vel_hist_px_dm_dist}, upper panel), and this is merely an artifact
of moving the MSPs too close to the Sun, reducing their apparent speeds.

We have performed a two-sample Kolmogorov-Smirnov test for comparing the continuous distributions of the parallax-based velocities and either of the two DM-based velocities to see if their macroscopic properties would
differ. For the YMW16 model, the maximum absolute differences between the cumulative distribution functions were $0.11$ and the p-value was $0.86$. For the NE2001 model, the maximum absolute differences between the cumulative distribution functions were $0.17$ and the p-values was $0.32$. There is thus no statistical evidence that the parallax-based distances and DM-based distances differ that markedly in $v_{\perp}$.

For the $35$ pulsars with significant parallaxes ($>3\sigma$), $\sim 2/3$ of their YMW16 DM-based distances are more than $1\sigma$ different from the parallax distances. This affects any velocities calculated from DM-based distances and makes the distribution of the velocities broader. It is beyond the scope of this current work to examine each DM model and search for reasons why they predict the wrong distances but using either the NE2001 or YMW16 models for calculating distances for finding 
velocities of individual objects is subject to large uncertainties and for
the rest of the paper we only consider the transverse velocities of
pulsars that have reliable parallax measurements. It is very difficult to calibrate
distance models at high galactic latitudes in the absence of pulsars with parallaxes.

\bsp	
\label{lastpage}
\end{document}